\documentclass[prl,aps,twocolumn,notitlepage,superscriptaddress,showpacs,nofootinbib]{revtex4-2}

\usepackage{enumerate,appendix}
\usepackage{amsmath, amsthm, commath,braket}
\usepackage{color,calc,graphicx}
\usepackage[usenames,dvipsnames,svgnames,table,cmyk,hyperref]{xcolor}
\usepackage[colorlinks]{hyperref}
\hypersetup{
	colorlinks = true,
	urlcolor = {blue},
	citecolor = {blue},
	linkcolor= {blue}
}

\usepackage{graphicx}
\usepackage{amsmath}
\usepackage{latexsym}
\usepackage{bbm}

\newcommand\mM{{\mathcal M }}
\newcommand\mN{{\mathcal N }}

\usepackage[charter,cal=cmcal,sfscaled=false]{mathdesign}
\usepackage{booktabs}
\usepackage{multirow}
\usepackage{subfigure}
\usepackage{dcolumn}

\begin{document}
\title{Testing Heisenberg's measurement uncertainty relation of three observables}

\author{Ya-Li Mao}

\affiliation{Shenzhen Institute for Quantum Science and Engineering and Department of Physics, Southern University of Science and Technology, Shenzhen, 518055, China}

\author{Hu Chen}
\affiliation{Shenzhen Institute for Quantum Science and Engineering and Department of Physics, Southern University of Science and Technology, Shenzhen, 518055, China}
\affiliation{Guangdong Provincial Key Laboratory of Quantum Science and Engineering, Southern University of Science and Technology, Shenzhen, 518055, China}

\author{Chang Niu}
\affiliation{Hefei National Laboratory for Physical Sciences at Microscale and Department of Modern Physics, University of Science and Technology of China, Hefei, Anhui 230026, China}

\author{Zheng-Da Li}

\affiliation{Shenzhen Institute for Quantum Science and Engineering and Department of Physics, Southern University of Science and Technology, Shenzhen, 518055, China}
\affiliation{Guangdong Provincial Key Laboratory of Quantum Science and Engineering, Southern University of Science and Technology, Shenzhen, 518055, China}

\author{Sixia Yu}

\affiliation{Hefei National Laboratory for Physical Sciences at Microscale and Department of Modern Physics, University of Science and Technology of China, Hefei, Anhui 230026, China}

\author{Jingyun Fan}

\affiliation{Shenzhen Institute for Quantum Science and Engineering and Department of Physics, Southern University of Science and Technology, Shenzhen, 518055, China}
\affiliation{Guangdong Provincial Key Laboratory of Quantum Science and Engineering, Southern University of Science and Technology, Shenzhen, 518055, China}
\affiliation{Center for Advanced Light Source, Southern University of Science and Technology, Shenzhen, 518055, China}

\begin{abstract}

Heisenberg's measurement uncertainty relations (MUR) of two quantum observables are essential for contemporary researches in quantum foundations and quantum information science. Going beyond, here we report the first experimental test of MURs for three quantum observables. Following the proposal of Bush, Lahti, and Werner [Phys. Rev. A 89, 012129 (2014)], we first establish rigorously MURs for triplets of unbiased qubit observables as combined approximation errors lower-bounded by an incompatibility measure. We then develop a convex programming protocol to numerically find the exact value of the incompatibility measure and the corresponding optimal measurements. Furthermore, we propose a novel implementation of optimal joint measurements and experimentally test our MURs using a single-photon qubit. Lastly, we discuss to analytically calculate the exact value of incompatibility measure for some symmetric triplets. We anticipate that this work may stimulate broad interests associated with the Heisenberg's uncertainty relation of multiple observables, enriching our understanding of quantum mechanics and inspiring innovative applications in quantum information science.
\end{abstract}

\maketitle 

\emph{Introduction ---}
Heisenberg's uncertainty principle~\cite{Heisenberg1927Physik} is one of the most distinctive features in which quantum meachanics differs from classical theories. The extensive exploration of the uncertainty principle about a pair of quantum observables has revealed two types of uncertainty relations, namely, the preparation uncertainty relations (PUR, also known as the Heisenberg-Robertson uncertainty relation)~\cite{Kennard1927Physik,Robertson1929PhysRev} and the measurement uncertainty relations (MUR) ~\cite{Ozawa2003PRA,Ozawa2004AOP,Ozawa2004PLA,Werner2004,BUSCH20061,Fujikawa2012PRA,Branciard2013PNAS,Busch2013PRL,Busch2014PRA,Yu2014arXiv,Busch2014RMP,PhysRevLett.114.070402,Bush2015arXiv,Coles2017RMP,math2018,yunger2019entropic}. While the PURs prohibit us from preparing quantum states with definite values for incompatible observables, the MURs capture the essence of quantum incompatibility~\cite{Heinosaari_2016,Guhne2021}, namely, quantum measurements may disturb each other, which was the main concern in the original Heisenberg's gedanken experiment of microscope~\cite{Heisenberg1927Physik}. These uncertainty relations deepen our understanding of quantum mechanics and crucially underlie quantum measurements and quantum information science~\cite{BUSCH20061,BUSCH2007155,RevModPhys.82.1155,Giovannetti2011nphoton,LIANG20111,RevModPhys.84.1655,Busch2014RMP,RevModPhys.91.025001,RevModPhys.94.025008}. Hence it is of high interest to explore uncertainty relations of multiple ($\ge3$) quantum observables from both fundamental and practical perspectives~\cite{PhysRev.46.794,Winter2008,PhysRevA.86.024101,math4010008,Maccone2014PRL,Kechrimparis2014PRA,Ma2017PRL,Fan2018PRA,LIANG20111,Yu2013arXiv,Schwonnek2016math4020038,Zhao2017PRA,Qin2019PRA,EP2022}. 

As some quantum measurements disturb each other, understanding the joint measurability of quantum measurements and achieving the optimal joint measurement with minimal error are critical tasks, which, however, had remained hard problems in quantum mechanics for decades. It was only until recent years that MURs of two observables have been conceived and experimentally verified
~\cite{Ozawa2003PRA,Ozawa2004AOP,Ozawa2004PLA,Lund2010NJP,Fujikawa2012PRA,Branciard2013PNAS,PhysRevLett.110.010404,Busch2013PRL,Busch2014PRA,Yu2014arXiv,Erhart2012Nphysics,Rozema2012PRL,Sulyok2013PRA,Baek2013SP,PhysRevLett.110.220402,Ringbauer2014PRL,Kaneda2014PRL,Ma2016PRL,Zhou2016SA,Xiong2017NJP,Zhao2017PRA,Bullock_2018,Mao2019PRL,yungerExp}.
For example, Bush, Lahti, and Werner (BLW) showed that we can approximate two incompatible measurements $(A,B)$ via the implementation of a pair of compatible, i.e., jointly measurable, quantum observables $(C, D)$ whose outcome can be read out simultaneously with errors. The resulting MUR indicates that the combined error quantified by statistic distances is lower bounded by the degree of incompatibility of $(A,B)$, which provides the performance characterization of the measuring device~\cite{Busch2013PRL, Busch2014PRA}. 
This elegant and operationally meaningful approach is immediately adapted to study the joint measurement of triple ideal qubit observables~\cite{Qin2019PRA,EP2022}. However, the optimal MUR with an exact value of the incompatibility and its experimental test remain elusive. 

In this contribution, following the proposal of BLW, we first establish the MUR for triplet of unbiased qubit observables by considering the most general jointly measurable triplet as approximation and provide a lower bound for the quantum incompatibility measure with exact condition of attainability.
Second, by framing the MUR as a convex optimization problem~\cite{Schwonnek2016math4020038}, we compile a convex programming protocol to numerically find the exact value of incompatibility measure and the corresponding optimal measurement. Based on a novel implementation of optimal joint measurement, we accomplish a few experimental demonstrations of saturated MURs. We stress that this is the first experimental test of MUR of multiple observables with an attainable lower bound, which is directly relevant to the fundamental limit of quantum precision measurement~\cite{PhysRevLett.114.070402,Xiong2017NJP}. Lastly, we present an analysis from the perspective of symmetry and illustrate to derive analytically the incompatibility measure for two symmetric triplets.

\emph{Basic theory ---} 
We briefly introduce the basic theory for the experimental realization of optimal joint measurement on three ideal qubit observables. Our target qubit measurements are three ideal qubit observables $\mM=\{{M}_i=\vec{m}_{i}\cdot\vec{\sigma}\}_{i=1}^3$, where $\vec{\sigma}=\{\sigma_{X},\sigma_{Y},\sigma_{Z}\}$ are Pauli matrices and $\vec m_i$ are unit Bloch vectors. If the qubit system is prepared in the state $\rho_s=({1+\vec{r}_s\cdot\vec{\sigma}})/{2}$, the distributions of the measurement outcomes are given by 
$P(\pm|M_i)= \frac{1\pm\vec m_i\cdot\vec r_s}2.$ The most general measurement of a qubit observable with two outcomes is described by the positive-operator-valued measures (POVM) $\mathcal M=\{ M^\pm\}$ with $ M^{\pm}=\frac{1\pm(x+{\vec{m}}\cdot\vec{\sigma})}{2}$, which is normalized, $M^++ M^-=1$, and nonnegative, $M^{\pm}\ge 0$, as long as $|x|+|\vec m|\le1$, where $\vec{m}$ is the Bloch vector and $\abs{x}$ stands for the biasedness.

A set of general qubit observables $\mathcal M$ are called jointly measurable if there exists a parent POVM $\mathcal R_p$ with multiple outcomes such that the each observable in the given set arises as a marginal measurement or equivalently from a post-measurement processing~\cite{Toigo2009}. The exact joint measurement conditions are known in a few special cases~\cite{PhysRevD.33.2253,Yu2010,Yu2013arXiv,PhysRevA2016} and in general the problem can be cast into semidefinite programming~\cite{PhysRevLett2009,Schneeloch2013PRA}. 
The necessary and sufficient conditions for three unbiased qubit observables to be compatible, i.e., with vanishing biasedness $x$, is given by~\cite{Yu2013arXiv}
\begin{equation}
	\sum_{k=0}^{3}|\vec{p}_k-\vec{p}_f|\le 4,
	\label{FMC}
\end{equation}
where $\vec p_k=\sum_{j=1}^3\gamma_{jk}\vec m_j$ with $\gamma_{jk}=(-1)^{k\lfloor\frac j2\rfloor+j \lfloor\frac k2\rfloor}$ and $\vec p_f$ is the Fermat-Torricelli (FT) point of $\{\vec{p}_k\}^3_{k=0}$, i.e., the vector that minimizes the left-hand-side in the above inequality.

The triplet $\mM$ of ideal qubit observables with unit Bloch vectors are unbiased and are not jointly measurable. In fact each pair of them are incompatible. Ideally we would like to have the information of each observable, $\{P(\pm|M_i)\}~$, somehow end up in a single measurement setup. Following BLW~\cite{Busch2013PRL, Busch2014PRA},
we may approximate these ideal measurements with three general observables $\mN=\{N_i\}$ that are jointly measurable. In general the actual measurement statistics $\{P(\pm|N_i)\}~$  deviate from the ideal measurement statistics $\{P(\pm|M_i)\}~$. Then the combined approximation errors are quantified by the total statistical distance $\Delta_\rho=\sum_{i=1}^3d_{\rho}(M_i; N_i)$ between the corresponding statistics, where
$d_{\rho}(M_i; N_i):= 2\sum_{\pm} |P(\pm|M_i) - P(\pm|N_i)|$ (see the illustration in Fig.1a). This leads to the state-independent definition of the measure of incompatibility 
\begin{equation}\label{imc}
	\Delta_\mM:=\min_{\mN}\max_\rho\Delta_\rho.
\end{equation}
This worst-case estimate of the inaccuracy characterizes the performance of the measurement device. 

An elegant lower bound of $\Delta_\mM$ was proposed in~\cite{Qin2019PRA} which is intimately related to the joint measurability condition. However, in deriving their result a strong presumption that the optimal measurements are unbiased was introduced. We strengthen this measurement uncertainty relation by considering the most general form of jointly measurable triplet by proving that  the optimal measurement is actually unbiased~\cite{ysx2022}.

{\bf Theorem 1} (Triplet MUR) For a triplet of ideal qubit observables $\mM=\{\vec{m}_{i}\cdot\vec{\sigma}\}_{i=1}^3$,
by performing the most general measurements that are jointly measurable, it holds MUR
\begin{equation}\label{MUR}
	\Delta_\mM\ge \frac{1}{2}\sum_{k=0}^{3}|\vec{p}_k-\vec{p}_f|-2:=2\delta
\end{equation}
where $\{\vec p_k=\sum_j\gamma_{jk}\vec m_j\}$ with $\vec p_f$ being its FT point. The lower bound is saturated if and only if
$\delta\le	\min_k|\vec p_k-\vec p_f|$. If the condition is met, the optimal set of jointly measurable triplet reads
$\vec n_{j}=\vec m_{j}+\frac\delta4\sum_{k=0}^3\gamma_{jk}\frac{\vec p_f-\vec p_{k}}{|\vec p_f-\vec p_{k}|}\ (k=1,2,3).$

We note that the MUR holds also for triplet of unbiased observables. As examples of attainability, the triplet with mutually orthogonal Bloch vectors, e.g., $\mM_o=\{\vec\sigma\cos\gamma\}$, can attain the equality, i.e., the MUR [Eq. (\ref{MUR})] is an optimal error trade-off relation for  $\gamma\le \arccos\frac1{\sqrt3}\approx 54.74^\circ$ for which $\mM_o$ is incompatible. In general, the MUR [Eq. (\ref{MUR})] cannot be attained for, e.g., coplanar triplet $\mM_p$ with degenerate FT point \cite{Yu2013arXiv}, i.e., $\vec p_f$ coincides with some $\vec p_k$. In these cases, the exact value of incompatibility $\Delta_\mM$ can be calculated via a convex programming.

\begin{figure*}
	\centering
	\includegraphics[width=\linewidth]{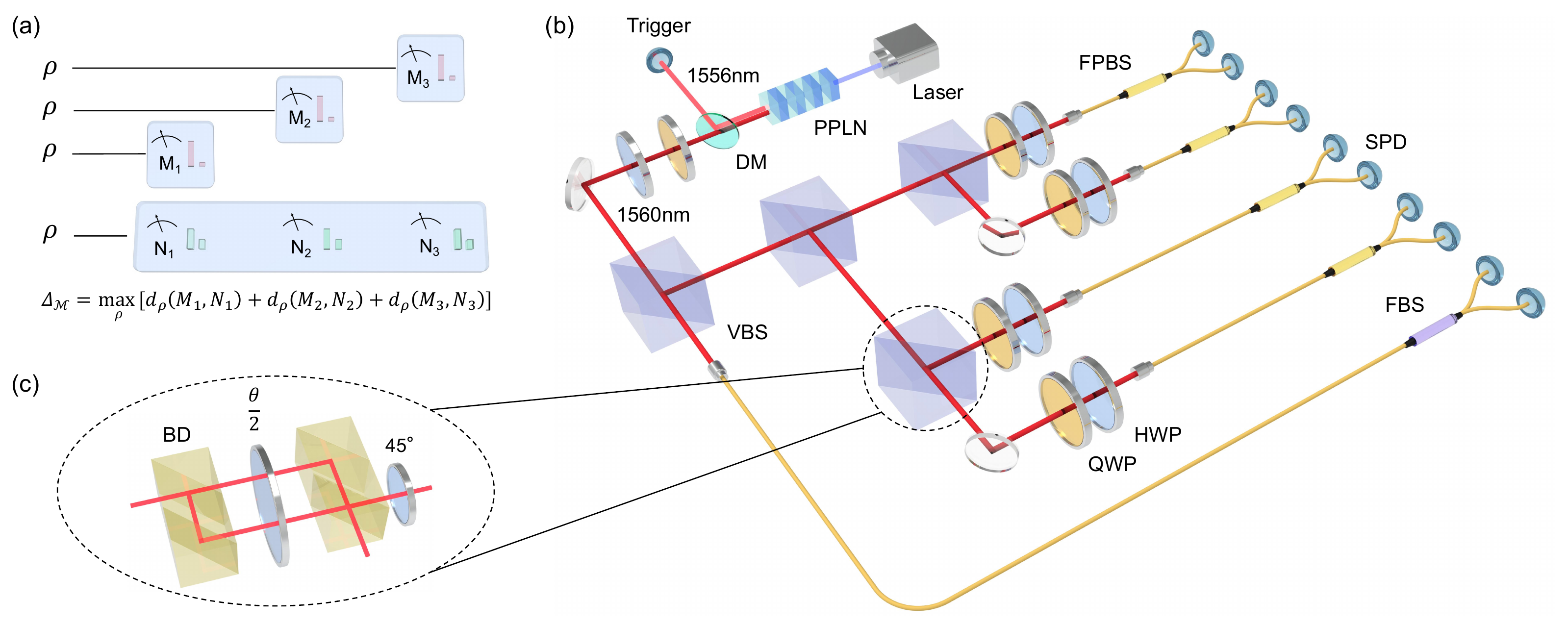}
	\caption{Experimental optimal joint measurement on triplets of qubit observables. (a) An illustration to realize the joint measurement on three jointly measurable observables $\{N_i\}$ to approximate that of three not jointly measurable observables $\{M_i\}$, where the combined approximation errors $\{\Delta_\rho(M_i,N_i)\}$ are maximized over all input quantum states $\rho_s$ and minimized over all triple jointly measurable observables. (b) A schamtics of experimental realization. We generate a pair of photons at the phase-matched wavelengths of 1556 nm (signal) and 1560 nm (idler) by passing a laser pulse at $\lambda_p=779$ nm through a periodically poled MgO doped Lithium Niobate (PPLN) crystal to induce the Type-0 spontaneous parametric down-conversion process~\cite{Li2022PRL}. The successful detection of a signal photon as a trigger signals the presence of an idler photon. We prepare the single qubit with the idler photon and implement the optimal joint measurement on the qubit by performing four polarization-projective measurements randomly and one identity measurement (see main text for details). (c) A realization of a variable beam splitter (VBS) with adjustable beam-splitting ratio.}
	\label{fig:Setup}
\end{figure*}

{\bf Protocol 1 } (Convex programming) The exact value of incompatibility $\Delta_{\mM}$ for a triplet $\{\vec m_j\}$ of ideal observables is given by the solution to the following convex optimization:
\begin{equation}
	\begin{aligned}
		\min_{{\mathcal R_p}=\{R_j\}}~~~&2\max_{k\in\{0,1,2,3\}}\abs{\textstyle\sum_{j=1}^3\gamma_{jk}(\vec m_j-\vec n_j)},\\
		\text{subj. to}~~~&  R_j \ge 0, ~~~(j=1,2,...,8),\ \sum_{j} R_j =I,\\
	\end{aligned}
	\label{theory112}
\end{equation}
with $\{\vec n_j\}$ being the Bloch vectors for three marginal measurements  $\{ N_{i}\}_{i=1}^3$ of $\mathcal R_p$.
Here, we use the CVX package under MATLAB, and mosek as CVX solver. 
The above protocol yields the exact value of incompatibility $\Delta_\mM$, the respective general measurement $\mathcal R_p$, and the optimal qubit state. As optimal measurement always lies on the boundary, i.e., saturating the joint measurement condition Eq. (\ref{FMC}), we may accomplish the optimal joint measurement in a single-qubit experiment~\cite{Yu2014arXiv}. 

{\bf Theorem 2} (Implementation) A jointly measurable triplet of unbiased qubit observables $\{ N_i\}$ that saturates the joint measurement condition Eq. (\ref{FMC}) can be implemented by the following parent measurement $\{R_{\mu_k|k}=P_k O_{\mu_k|k}\}$, where
\begin{equation}\label{Implementation}
	P_k=\frac{|\vec q_k-\vec q_f|}4,\quad O_{\mu_k|k}= \frac12\left(1+\mu_k\frac{\vec q_k-\vec q_f}{|\vec q_k-\vec q_f|}\cdot\vec\sigma \right), 
\end{equation}
with binary outcomes labeled with $\mu_k=\pm1$ for each $k=0,1,2,3$, and $\vec q_f$ is the FT point of $\{\vec{q}_k\}$. Hence the marginal measurements are ${N}_{\mu|j}=\sum_{k,\mu_k}p_j(\mu|k,\mu_k)R_{\mu_k|k}$  with post-measurement processing $p_j(\mu|k,\mu_k)=\frac{1+\mu\ \gamma_{jk}\ \mu_k}2$,  where $j=1,2,3$ and binary outcomes are labeled with $\mu=\pm1$.

\emph{Experiment---} Here we present an experimental demonstration of implementing the optimal joint measurement on ideal triple qubit observables with a photonic polarization qubit, with the experimental schematics depicted in Fig. \ref{fig:Setup}, following Theorem 2. 

By passing a laser pulse with wavelength at $\lambda_p=779$ nm through a piece of periodically poled MgO doped Lithium Niobate (PPLN) crystal to induce Type-0 spontaneous parametric down-conversion (SPDC) process, we probabilistically create a pair of photons at the phase-matched wavelengths of 1560 nm (signal) and 1556 nm (idler)~\cite{Li2022PRL}. The detection of the idler photon heralds the presence of a signal photon. 
With a pair of half- and quarter- waveplates (HWP, QWP), we can prepare the qubit, here the polarization state of the photon, $\ket{\Phi}_{s}=\cos\alpha\ket{H}_{s}+e^{i\phi}\sin\alpha\ket{V}_{s}$, arbitrarily on the Bloch sphere, where $\alpha/2$ is the angle of the fast axis of a HWP oriented from the vertical, $\phi$ is the phase, and $\ket{H}_{s}$ and $\ket{V}_{s}$ stand for horizontal and vertical polarization states, respectively. 

We configure a group of beam splitters to direct the single photon to one identity detection module and four polarization-detection modules, respectively. The identity module consists of a 50:50 fiber beam splitter (FBS) and each polarization detection-module consists of a pair of QWP and HWP and a fiber polarizing beam splitter (FPBS) to implement the polarization-projection measurement on the photonic qubit.

We conduct the experiment following the procedure: For a triplet $\{{M}_i\}$, we follow Protocol 1 to find the respective general measurement $\mathcal R_p$ and the qubit state, and follow Theorem 2 to design the optimal measurements $\{O_k\}$. After that, we prepare the single-photon qubit. We first make the beam splitter group and the detection modules to perform the polarization-projection measurements $\{{M}^{\pm}_i\}$, then reset the splitting ratios of beam splitters and the detection modules to perform the polarization-projection measurements $\{{O}_{{\pm}|k}\}$ with probabilities $\{P_k\}$ according to Eq. (\ref{Implementation}) to accomplish the optimal joint measurement $\{ N_{i}\}_{i=1}^3$. The two sets of measurement statistics yield the lower bound of incompatibility $\Delta_{\mM}$ that we can attain in the experiment.

\begin{figure}
	\centering
	\includegraphics[width=1.\linewidth]{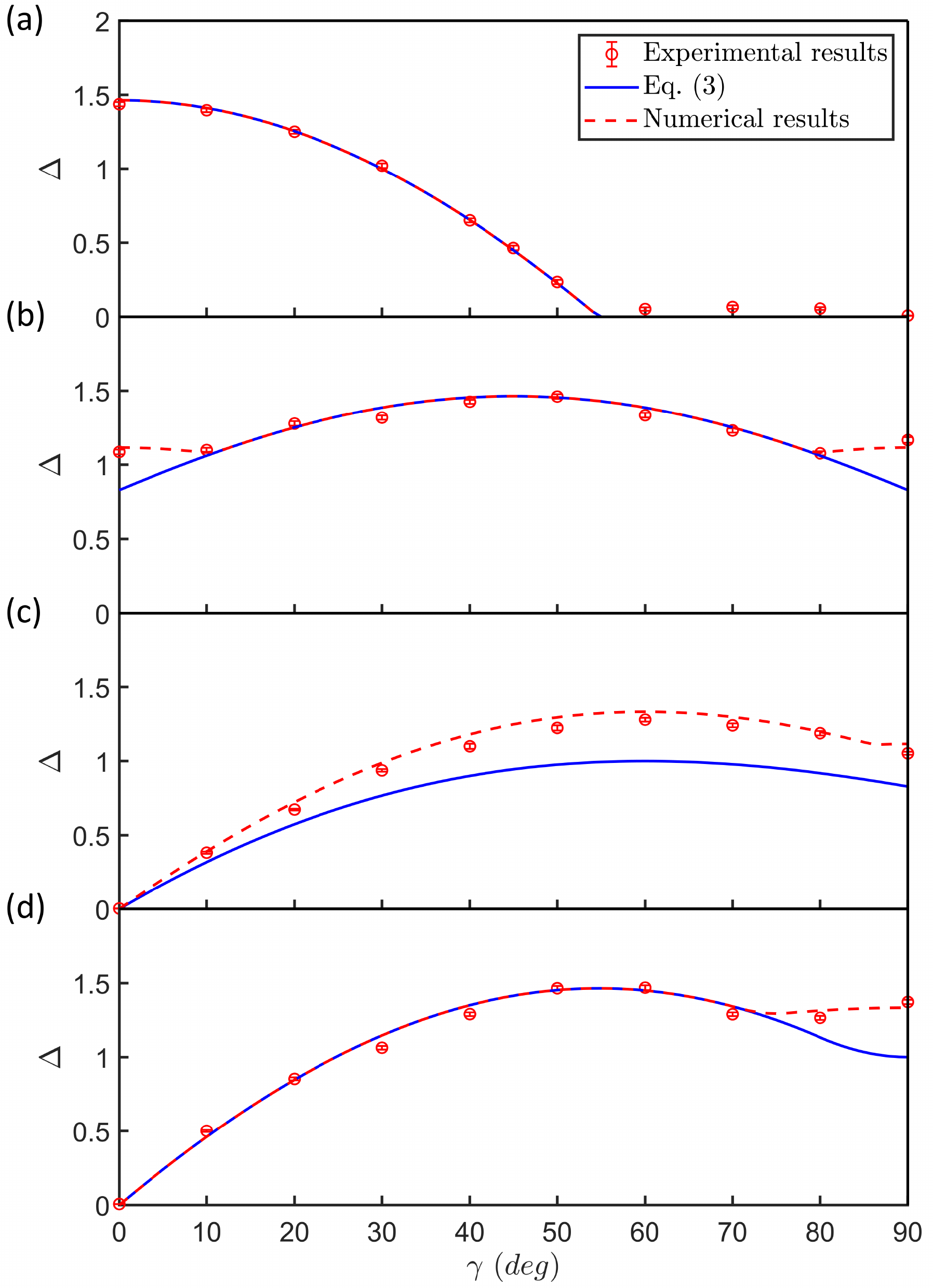}
	\caption{Exact value of incompatibility of 4 triplets of idea qubit observables: 
	(a)$\mM_o$, 
	(b)$\mM_\perp$,
	(c)$\mM_p$,
	(d)$\mM_Y$. 
	Red dashed lines are numerical results calculated with Protocol 1, blue smooth curves are lower bound of incompatibility in Eq. (\ref{MUR}), and red triangles represent experimental results. Error bars stand for one-standard deviations.}
	\label{fig:F-3O}
\end{figure}

We investigate the optimal joint-measurement on triple ideal qubit observables for a few selected scenarios, (i) triplet $\mM_o=\{\sigma_{Z},\sigma_{Y},\sigma_{X}\}\cos\gamma$ with Bloch vectors pairwise orthogonal; (ii) triplet $\mM_\perp=\{\sigma_{X}\cos\gamma+\sigma_{Y}\sin\gamma,\sigma_{X}\cos\gamma+\sigma_{Y}\sin\gamma,\sigma_{Z}\}$ with one Bloch vector orthogonal to the plane spanned by the other two; (iii) co-planar triplet $\mM_p=\{\sigma_{X}\cos\gamma+\sigma_{Y}\sin\gamma, \sigma_{X}\cos\gamma-\sigma_{Y}\sin\gamma,\sigma_{X}\}$; and (iv) triplet $\mM_Y=\{\frac{-\sigma_{X}+\sqrt{3}\sigma_{Y}}2,\frac{-\sigma_{X}-\sqrt{3}\sigma_{Y}}2,\sigma_{X}\}\sin\gamma+\sigma_{Z}\cos\gamma$ with Bloch vectors being neither orthogonal and nor pairwise co-planar. The results are plotted respectively in Fig. \ref{fig:F-3O}(a-d), with angle parameter $\gamma\in [0^\circ,90^\circ]$.  We draw the lower bounds on the right hand side (RHS) of Eq. (\ref{MUR}) with blue smooth lines, the attainable lower bounds obtained from the convex optimizing program (Protocol 1) with red dashed lines, and experimental results with open dots. 

Some remarks are in order. First, comparing the red dashed lines and blue lines, it is evident that the MUR of Eq. (\ref{MUR}) is optimal, i.e., with attainable lower bound, for the entire parameter range of $\gamma$ in scenario (i) and for parts of the parameter range in (ii) and (iv), and not optimal, i.e., with RHS smaller than that of the attainable lower bound found by Protocol 1, for the other parts of the parameter range in (ii) and (iv) and the entire parameter range in (iii). We note that the region of $\gamma$ that attains the MUR can be determined by Theorem 1. Second, experimental results are consistently in good agreement with numerical results obtained via Protocol 1 for all scenarios under study, i.e., we experimentally attain the exact value of incompatibility $\Delta_{\mM}$. This justifies the strategy of accomplishing the optimal joint measurement with single-qubit given in Theorem 2.

\emph{Discussion ---}
Symmetry is considered to be the main characteristics of the laws of physics according to Feynman~\cite{Feynman} and plays an essential role in determining incompatibility noise robustness~\cite{Toigo2018,Guhne2020}.
For some physically relevant sets of measurement there are certain symmetry among measurement directions. For some physically relevant sets of measurement there are certain symmetry among measurement directions. For example in the case of an ideal triplet $\mM_\perp$ with one observable ($\vec m_3$ along $\vec{z}$ direction) being orthogonal to the other two ($\vec m_{1,2}$ on the $XY$ plane) the reflection $\tau_{XY}$ over $XY$ plane and  the reflections $\tau_{\pm}$ over planes passing through $\vec z$ and angle bisectors $\vec m_1\pm\vec m_2$ of $\vec m_{1,2}$ are a kind of graded symmetry. By a graded symmetry here we mean a reflection such that three directions are permuted among themselves up to some inversions (which correspond to relabeling two outcomes of a measurement). For example, we have
$\tau_{XY}\cdot \vec m_3=-\vec m_3$ while preserving directions $\vec m_{1,2}$ and $\tau_\pm\cdot\vec m_{k}=\pm\vec m_{2-k}$ with $k=1,2$ while preserving $\vec m_3$.  It turns out that the optimal measurement must share the same graded symmetry.

\begin{figure}
	\includegraphics[width=1.\linewidth]{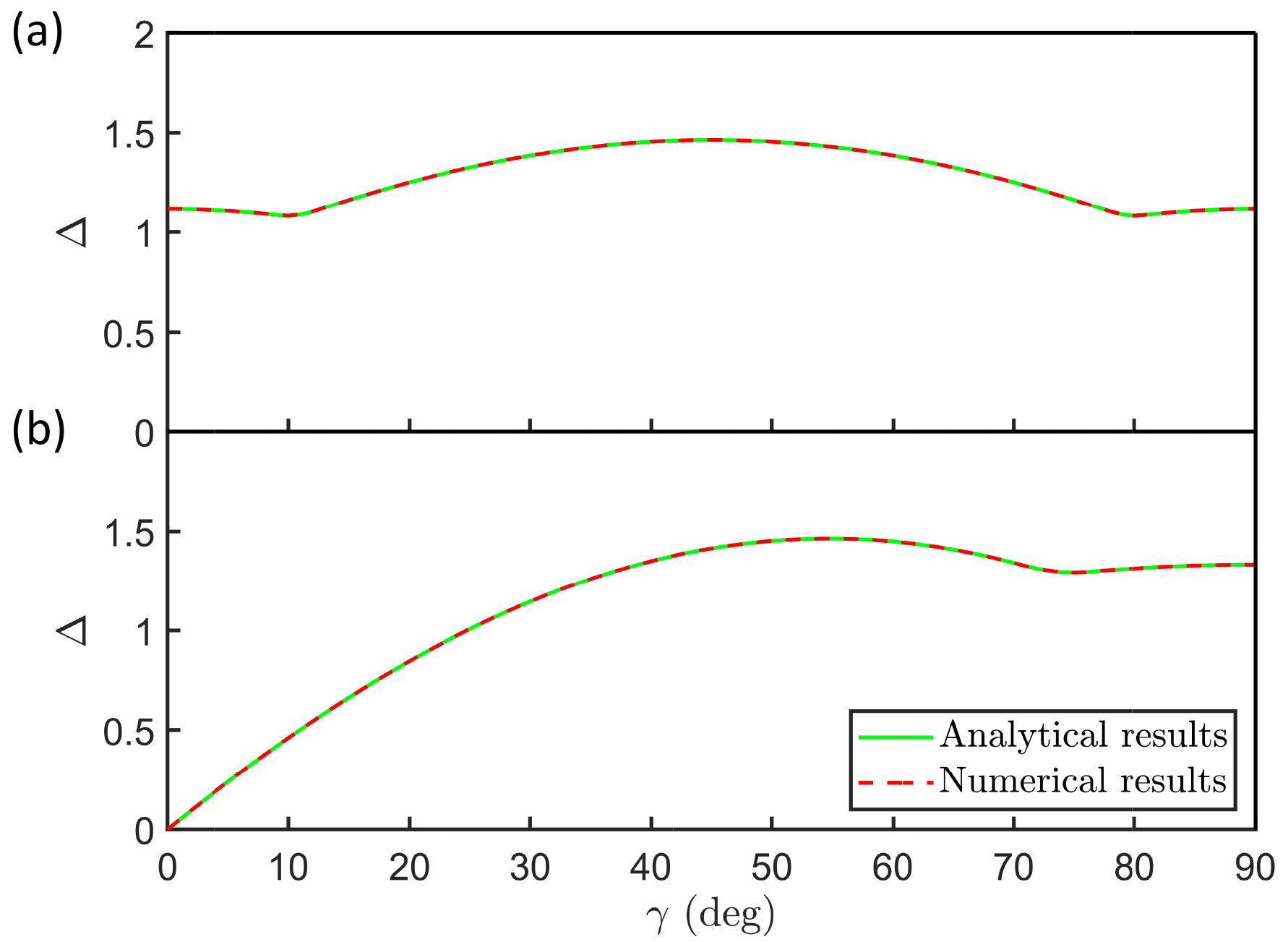}
	\caption{Exact values of incompatibility for two triplets of ideal qubit measurements with symmetries, (a) triplet $\mM_\perp$ with $|\cos2\gamma|=\vec m_1\cdot\vec m_2$ and (b) highly symmetric triplet $\mM_Y$ of unbiased observables (see main text for details). Red dashed lines are numerical results found with Protocol 1 and green lines are analytical results computed with Eqs. (\ref{sym1}) and (\ref{sym2}), respectively. }
	\label{fig:F-sym}
\end{figure}

{\bf Theorem 3} (Symmetry) If the ideal triplet $\mM$ along directions $\{\vec m_j\}_{j=1}^3$ admits a graded symmetry $g$, i.e., a reflection such that
$g\cdot \vec m_j={\omega_j} \vec m_{\sigma(j)}$
with  $\omega_j=\pm$ and $\sigma$ being a permutation of $\{1,2,3\}$, then the optimal jointly measurable triplet $\mN=\{\vec n_j\cdot\vec{\sigma}\}_{j=1}^3$ for the incompatibility $\Delta_\mM$ shares the same graded symmetry, i.e.,
$g\cdot \vec n_j={\omega_j}  \vec n_{\sigma(j)}.$

For some sets of ideal observables their symmetry might determine the sets itself. For example, the triplet $\mM_o=\{\sigma_k\}$ of three orthogonal observables is completely determined by 3 reflections over planes $XY,YZ,ZX$ upto some scalings. Thus the optimal joint measurements are along the same directions and the incompatibility is readily found to be $\Delta_o=2(\sqrt 3\cos\gamma-1)$. For some sets their symmetry might determine the set partially (e.g. the symmetry of $\mM_\perp$). Thus this inheriting symmetry enables us to calculate analytically some symmetric triplet of ideal observables. For triplet $\mM_\perp$ with $\abs{\cos2\gamma}=\vec m_1\cdot\vec m_2$ the incompatibility reads%
\begin{equation}\label{sym1}
	\Delta_{\perp}=\left\{\begin{array}{ll}2{\sqrt{2+\sin2\gamma}-2} & |\gamma-45^\circ|\le\Gamma_0\\
		{2\sqrt{3+\sin2\gamma-2\sqrt{\sin2\gamma}-2\abs{\cos2\gamma}}}& \mbox{otherwise}\\
		\Delta_{opt}(\gamma)& |\gamma-45^\circ|\ge\Gamma_1\\
	\end{array}\right.
\end{equation}
where $\Delta_{opt}=(1-\cos t)\sqrt{\sec^2t+3}$ with $t$ determined by
$\abs{\cos2\gamma}=\frac18\tan^2t(1+3\cos t)^2-1$ and $\Gamma_0\approx 32.77^\circ$ determined by the attainability criterion in Theorem 1, and $\Gamma_1\approx 35.77^\circ$. The corresponding optimal measurements for different regions of $\gamma$ can also be analytically given\cite{ysx2022}. 

Our second example is  the following highly symmetric triplet $\mM_Y$ of unbiased observables along directions
$\vec m_j=\vec{z}\cos\gamma+\vec e_j\sin\gamma,$ $\vec e_j\cdot\vec e_k=-\frac 12,\ \vec z\cdot\vec e_j=0, \ 0\le\gamma\le\frac\pi2$, where $\vec e_j,\vec e_k$ are unit vectors.
For the triplet $\mM_Y$, we have the exact incompatibility
\begin{equation}\label{sym2}
	\Delta_{Y}=\left\{\begin{array}{ll}{2\cos\gamma+2\sqrt 2\sin\gamma-2} \\
		{\sqrt2{\sin \gamma}+4\cos\gamma- 2\sqrt{\frac{2}{3}-\left(\sin\gamma-\sqrt{2} \cos\gamma \right)^2}}\\
		{\displaystyle\min_{x^2+y^2=\frac19}2\sqrt{(\cos\gamma-x)^2+4(\sin\gamma-2y)^2}}
	\end{array}\right.
\end{equation}
for three intervals $0<\gamma_0<\gamma_1<90^\circ$ divided by $\gamma_0\approx 70.53^\circ$ and $\gamma_1\approx75.80^\circ$.

We plot the incompatibility (thick green lines) computed with Eqs. (\ref{sym1}) and (\ref{sym2}) in Fig. \ref{fig:F-sym} which perfectly coincide with numerical results (red dashed lines) obtained by convex programming Protocol 1.

\emph{Summary ---}
Some quantum measurements disturb each other, preventing us from measuring them  with a single measurement device without introducing errors. The MUR sets the limit to how well we can perform the joint measurement with the minimal amount of errors according to quantum mechanics.  
Following the proposal of BLW, we establish the MUR for triplets of qubit observables. Employing the convex programming, we find the exact value of the incompatibility measure and the optimal measurements to saturate the MUR. This guides us to accomplish the optimal joint measurements on triplets of qubit observables for the first time. As a critical step in the study of joint-measurement on multiple observables, this study may deepen our understanding of Heisenberg's measurement uncertainty principle and lead to new applications in quantum metrology and quantum information science. We stress that the demonstrated strategy can be readily generalized to calculate the incompatibility of four or more qubit observables and the case of weighted measurements, which will be considered elsewhere.

This work is supported by the Key-Area Research and Development Program of Guangdong Province Grant No.2020B0303010001, Grant No.2019ZT08X324, National Natural Science Foundation of China Grants No.12004207 and No.12005090, Shenzhen Science and Technology Program Grant No.RCYX20210706092043065, Guangdong Provincial Key Laboratory Grant No.2019B121203002 and SIQSE202104. 

\bibliography{Ref-3O}

\begin{thebibliography}{67}%
\makeatletter
\providecommand \@ifxundefined [1]{%
 \@ifx{#1\undefined}
}%
\providecommand \@ifnum [1]{%
 \ifnum #1\expandafter \@firstoftwo
 \else \expandafter \@secondoftwo
 \fi
}%
\providecommand \@ifx [1]{%
 \ifx #1\expandafter \@firstoftwo
 \else \expandafter \@secondoftwo
 \fi
}%
\providecommand \natexlab [1]{#1}%
\providecommand \enquote  [1]{``#1''}%
\providecommand \bibnamefont  [1]{#1}%
\providecommand \bibfnamefont [1]{#1}%
\providecommand \citenamefont [1]{#1}%
\providecommand \href@noop [0]{\@secondoftwo}%
\providecommand \href [0]{\begingroup \@sanitize@url \@href}%
\providecommand \@href[1]{\@@startlink{#1}\@@href}%
\providecommand \@@href[1]{\endgroup#1\@@endlink}%
\providecommand \@sanitize@url [0]{\catcode `\\12\catcode `\$12\catcode
  `\&12\catcode `\#12\catcode `\^12\catcode `\_12\catcode `\%12\relax}%
\providecommand \@@startlink[1]{}%
\providecommand \@@endlink[0]{}%
\providecommand \url  [0]{\begingroup\@sanitize@url \@url }%
\providecommand \@url [1]{\endgroup\@href {#1}{\urlprefix }}%
\providecommand \urlprefix  [0]{URL }%
\providecommand \Eprint [0]{\href }%
\providecommand \doibase [0]{https://doi.org/}%
\providecommand \selectlanguage [0]{\@gobble}%
\providecommand \bibinfo  [0]{\@secondoftwo}%
\providecommand \bibfield  [0]{\@secondoftwo}%
\providecommand \translation [1]{[#1]}%
\providecommand \BibitemOpen [0]{}%
\providecommand \bibitemStop [0]{}%
\providecommand \bibitemNoStop [0]{.\EOS\space}%
\providecommand \EOS [0]{\spacefactor3000\relax}%
\providecommand \BibitemShut  [1]{\csname bibitem#1\endcsname}%
\let\auto@bib@innerbib\@empty
\bibitem [{\citenamefont {Heisenberg}(1927)}]{Heisenberg1927Physik}%
  \BibitemOpen
  \bibfield  {author} {\bibinfo {author} {\bibfnamefont {W.}~\bibnamefont
  {Heisenberg}},\ }\bibfield  {title} {\bibinfo {title} {{\"U}ber den
  anschaulichen inhalt der quantentheoretischen kinematik und mechanik},\
  }\href {https://doi.org/10.1007/BF01397280} {\bibfield  {journal} {\bibinfo
  {journal} {Zeitschrift f{\"u}r Physik}\ }\textbf {\bibinfo {volume} {43}},\
  \bibinfo {pages} {172} (\bibinfo {year} {1927})}\BibitemShut {NoStop}%
\bibitem [{\citenamefont {Kennard}(1927)}]{Kennard1927Physik}%
  \BibitemOpen
  \bibfield  {author} {\bibinfo {author} {\bibfnamefont {E.~H.}\ \bibnamefont
  {Kennard}},\ }\bibfield  {title} {\bibinfo {title} {Zur quantenmechanik
  einfacher bewegungstypen},\ }\href {https://doi.org/10.1007/BF01391200}
  {\bibfield  {journal} {\bibinfo  {journal} {Zeitschrift f{\"u}r Physik}\
  }\textbf {\bibinfo {volume} {44}},\ \bibinfo {pages} {326} (\bibinfo {year}
  {1927})}\BibitemShut {NoStop}%
\bibitem [{\citenamefont {Robertson}(1929)}]{Robertson1929PhysRev}%
  \BibitemOpen
  \bibfield  {author} {\bibinfo {author} {\bibfnamefont {H.~P.}\ \bibnamefont
  {Robertson}},\ }\bibfield  {title} {\bibinfo {title} {The uncertainty
  principle},\ }\href {https://doi.org/10.1103/PhysRev.34.163} {\bibfield
  {journal} {\bibinfo  {journal} {Phys. Rev.}\ }\textbf {\bibinfo {volume}
  {34}},\ \bibinfo {pages} {163} (\bibinfo {year} {1929})}\BibitemShut
  {NoStop}%
\bibitem [{\citenamefont {Ozawa}(2003)}]{Ozawa2003PRA}%
  \BibitemOpen
  \bibfield  {author} {\bibinfo {author} {\bibfnamefont {M.}~\bibnamefont
  {Ozawa}},\ }\bibfield  {title} {\bibinfo {title} {Universally valid
  reformulation of the {H}eisenberg uncertainty principle on noise and
  disturbance in measurement},\ }\href
  {https://doi.org/10.1103/PhysRevA.67.042105} {\bibfield  {journal} {\bibinfo
  {journal} {Phys. Rev. A}\ }\textbf {\bibinfo {volume} {67}},\ \bibinfo
  {pages} {042105} (\bibinfo {year} {2003})}\BibitemShut {NoStop}%
\bibitem [{\citenamefont {Ozawa}(2004{\natexlab{a}})}]{Ozawa2004AOP}%
  \BibitemOpen
  \bibfield  {author} {\bibinfo {author} {\bibfnamefont {M.}~\bibnamefont
  {Ozawa}},\ }\bibfield  {title} {\bibinfo {title} {Uncertainty relations for
  noise and disturbance in generalized quantum measurements},\ }\href
  {https://doi.org/https://doi.org/10.1016/j.aop.2003.12.012} {\bibfield
  {journal} {\bibinfo  {journal} {Annals of Physics}\ }\textbf {\bibinfo
  {volume} {311}},\ \bibinfo {pages} {350 } (\bibinfo {year}
  {2004}{\natexlab{a}})}\BibitemShut {NoStop}%
\bibitem [{\citenamefont {Ozawa}(2004{\natexlab{b}})}]{Ozawa2004PLA}%
  \BibitemOpen
  \bibfield  {author} {\bibinfo {author} {\bibfnamefont {M.}~\bibnamefont
  {Ozawa}},\ }\bibfield  {title} {\bibinfo {title} {Uncertainty relations for
  joint measurements of noncommuting observables},\ }\href
  {https://doi.org/https://doi.org/10.1016/j.physleta.2003.12.001} {\bibfield
  {journal} {\bibinfo  {journal} {Physics Letters A}\ }\textbf {\bibinfo
  {volume} {320}},\ \bibinfo {pages} {367 } (\bibinfo {year}
  {2004}{\natexlab{b}})}\BibitemShut {NoStop}%
\bibitem [{\citenamefont {{Werner}}(2004)}]{Werner2004}%
  \BibitemOpen
  \bibfield  {author} {\bibinfo {author} {\bibfnamefont {R.~F.}\ \bibnamefont
  {{Werner}}},\ }\bibfield  {title} {\bibinfo {title} {{The uncertainty
  relation for joint measurement of position and momentum}},\ }\href
  {https://arxiv.org/abs/quant-ph/0405184} {\bibfield  {journal} {\bibinfo
  {journal} {arXiv preprint quant-ph/0405184}\ } (\bibinfo {year}
  {2004})}\BibitemShut {NoStop}%
\bibitem [{\citenamefont {Busch}\ and\ \citenamefont
  {Shilladay}(2006)}]{BUSCH20061}%
  \BibitemOpen
  \bibfield  {author} {\bibinfo {author} {\bibfnamefont {P.}~\bibnamefont
  {Busch}}\ and\ \bibinfo {author} {\bibfnamefont {C.}~\bibnamefont
  {Shilladay}},\ }\bibfield  {title} {\bibinfo {title} {Complementarity and
  uncertainty in {M}ach–{Z}ehnder interferometry and beyond},\ }\href
  {https://doi.org/https://doi.org/10.1016/j.physrep.2006.09.001} {\bibfield
  {journal} {\bibinfo  {journal} {Physics Reports}\ }\textbf {\bibinfo {volume}
  {435}},\ \bibinfo {pages} {1} (\bibinfo {year} {2006})}\BibitemShut {NoStop}%
\bibitem [{\citenamefont {Fujikawa}(2012)}]{Fujikawa2012PRA}%
  \BibitemOpen
  \bibfield  {author} {\bibinfo {author} {\bibfnamefont {K.}~\bibnamefont
  {Fujikawa}},\ }\bibfield  {title} {\bibinfo {title} {Universally valid
  {H}eisenberg uncertainty relation},\ }\href
  {https://doi.org/10.1103/PhysRevA.85.062117} {\bibfield  {journal} {\bibinfo
  {journal} {Phys. Rev. A}\ }\textbf {\bibinfo {volume} {85}},\ \bibinfo
  {pages} {062117} (\bibinfo {year} {2012})}\BibitemShut {NoStop}%
\bibitem [{\citenamefont {Branciard}(2013)}]{Branciard2013PNAS}%
  \BibitemOpen
  \bibfield  {author} {\bibinfo {author} {\bibfnamefont {C.}~\bibnamefont
  {Branciard}},\ }\bibfield  {title} {\bibinfo {title} {Error-tradeoff and
  error-disturbance relations for incompatible quantum measurements},\ }\href
  {https://doi.org/10.1073/pnas.1219331110} {\bibfield  {journal} {\bibinfo
  {journal} {Proceedings of the National Academy of Sciences}\ }\textbf
  {\bibinfo {volume} {110}},\ \bibinfo {pages} {6742} (\bibinfo {year}
  {2013})}\BibitemShut {NoStop}%
\bibitem [{\citenamefont {Busch}\ \emph {et~al.}(2013)\citenamefont {Busch},
  \citenamefont {Lahti},\ and\ \citenamefont {Werner}}]{Busch2013PRL}%
  \BibitemOpen
  \bibfield  {author} {\bibinfo {author} {\bibfnamefont {P.}~\bibnamefont
  {Busch}}, \bibinfo {author} {\bibfnamefont {P.}~\bibnamefont {Lahti}},\ and\
  \bibinfo {author} {\bibfnamefont {R.~F.}\ \bibnamefont {Werner}},\ }\bibfield
   {title} {\bibinfo {title} {Proof of {H}eisenberg's error-disturbance
  relation},\ }\href {https://doi.org/10.1103/PhysRevLett.111.160405}
  {\bibfield  {journal} {\bibinfo  {journal} {Phys. Rev. Lett.}\ }\textbf
  {\bibinfo {volume} {111}},\ \bibinfo {pages} {160405} (\bibinfo {year}
  {2013})}\BibitemShut {NoStop}%
\bibitem [{\citenamefont {Busch}\ \emph
  {et~al.}(2014{\natexlab{a}})\citenamefont {Busch}, \citenamefont {Lahti},\
  and\ \citenamefont {Werner}}]{Busch2014PRA}%
  \BibitemOpen
  \bibfield  {author} {\bibinfo {author} {\bibfnamefont {P.}~\bibnamefont
  {Busch}}, \bibinfo {author} {\bibfnamefont {P.}~\bibnamefont {Lahti}},\ and\
  \bibinfo {author} {\bibfnamefont {R.~F.}\ \bibnamefont {Werner}},\ }\bibfield
   {title} {\bibinfo {title} {Heisenberg uncertainty for qubit measurements},\
  }\href {https://doi.org/10.1103/PhysRevA.89.012129} {\bibfield  {journal}
  {\bibinfo  {journal} {Phys. Rev. A}\ }\textbf {\bibinfo {volume} {89}},\
  \bibinfo {pages} {012129} (\bibinfo {year} {2014}{\natexlab{a}})}\BibitemShut
  {NoStop}%
\bibitem [{\citenamefont {Yu}\ and\ \citenamefont {Oh}(2014)}]{Yu2014arXiv}%
  \BibitemOpen
  \bibfield  {author} {\bibinfo {author} {\bibfnamefont {S.}~\bibnamefont
  {Yu}}\ and\ \bibinfo {author} {\bibfnamefont {C.}~\bibnamefont {Oh}},\
  }\bibfield  {title} {\bibinfo {title} {Optimal joint measurement of two
  observables of a qubit},\ }\href {https://arxiv.org/abs/1402.3785} {\bibfield
   {journal} {\bibinfo  {journal} {arXiv preprint arXiv:1402.3785}\ } (\bibinfo
  {year} {2014})}\BibitemShut {NoStop}%
\bibitem [{\citenamefont {Busch}\ \emph
  {et~al.}(2014{\natexlab{b}})\citenamefont {Busch}, \citenamefont {Lahti},\
  and\ \citenamefont {Werner}}]{Busch2014RMP}%
  \BibitemOpen
  \bibfield  {author} {\bibinfo {author} {\bibfnamefont {P.}~\bibnamefont
  {Busch}}, \bibinfo {author} {\bibfnamefont {P.}~\bibnamefont {Lahti}},\ and\
  \bibinfo {author} {\bibfnamefont {R.~F.}\ \bibnamefont {Werner}},\ }\bibfield
   {title} {\bibinfo {title} {Colloquium: Quantum root-mean-square error and
  measurement uncertainty relations},\ }\href
  {https://doi.org/10.1103/RevModPhys.86.1261} {\bibfield  {journal} {\bibinfo
  {journal} {Rev. Mod. Phys.}\ }\textbf {\bibinfo {volume} {86}},\ \bibinfo
  {pages} {1261} (\bibinfo {year} {2014}{\natexlab{b}})}\BibitemShut {NoStop}%
\bibitem [{\citenamefont {Busch}\ and\ \citenamefont
  {Stevens}(2015)}]{PhysRevLett.114.070402}%
  \BibitemOpen
  \bibfield  {author} {\bibinfo {author} {\bibfnamefont {P.}~\bibnamefont
  {Busch}}\ and\ \bibinfo {author} {\bibfnamefont {N.}~\bibnamefont
  {Stevens}},\ }\bibfield  {title} {\bibinfo {title} {Direct tests of
  measurement uncertainty relations: What it takes},\ }\href
  {https://doi.org/10.1103/PhysRevLett.114.070402} {\bibfield  {journal}
  {\bibinfo  {journal} {Phys. Rev. Lett.}\ }\textbf {\bibinfo {volume} {114}},\
  \bibinfo {pages} {070402} (\bibinfo {year} {2015})}\BibitemShut {NoStop}%
\bibitem [{\citenamefont {Bullock}\ and\ \citenamefont
  {Busch}(2018{\natexlab{a}})}]{Bush2015arXiv}%
  \BibitemOpen
  \bibfield  {author} {\bibinfo {author} {\bibfnamefont {T.}~\bibnamefont
  {Bullock}}\ and\ \bibinfo {author} {\bibfnamefont {P.}~\bibnamefont
  {Busch}},\ }\bibfield  {title} {\bibinfo {title} {Measurement uncertainty
  relations: characterising optimal error bounds for qubits},\ }\href
  {https://doi.org/10.1088/1751-8121/aac729} {\bibfield  {journal} {\bibinfo
  {journal} {Journal of Physics A: Mathematical and Theoretical}\ }\textbf
  {\bibinfo {volume} {51}},\ \bibinfo {pages} {283001} (\bibinfo {year}
  {2018}{\natexlab{a}})}\BibitemShut {NoStop}%
\bibitem [{\citenamefont {Coles}\ \emph {et~al.}(2017)\citenamefont {Coles},
  \citenamefont {Berta}, \citenamefont {Tomamichel},\ and\ \citenamefont
  {Wehner}}]{Coles2017RMP}%
  \BibitemOpen
  \bibfield  {author} {\bibinfo {author} {\bibfnamefont {P.~J.}\ \bibnamefont
  {Coles}}, \bibinfo {author} {\bibfnamefont {M.}~\bibnamefont {Berta}},
  \bibinfo {author} {\bibfnamefont {M.}~\bibnamefont {Tomamichel}},\ and\
  \bibinfo {author} {\bibfnamefont {S.}~\bibnamefont {Wehner}},\ }\bibfield
  {title} {\bibinfo {title} {Entropic uncertainty relations and their
  applications},\ }\href {https://doi.org/10.1103/RevModPhys.89.015002}
  {\bibfield  {journal} {\bibinfo  {journal} {Rev. Mod. Phys.}\ }\textbf
  {\bibinfo {volume} {89}},\ \bibinfo {pages} {015002} (\bibinfo {year}
  {2017})}\BibitemShut {NoStop}%
\bibitem [{\citenamefont {Barchielli}\ \emph
  {et~al.}(2018{\natexlab{a}})\citenamefont {Barchielli}, \citenamefont
  {Gregoratti},\ and\ \citenamefont {Toigo}}]{math2018}%
  \BibitemOpen
  \bibfield  {author} {\bibinfo {author} {\bibfnamefont {A.}~\bibnamefont
  {Barchielli}}, \bibinfo {author} {\bibfnamefont {M.}~\bibnamefont
  {Gregoratti}},\ and\ \bibinfo {author} {\bibfnamefont {A.}~\bibnamefont
  {Toigo}},\ }\bibfield  {title} {\bibinfo {title} {Measurement uncertainty
  relations for discrete observables: Relative entropy formulation},\ }\href
  {https://doi.org/https://doi.org/10.1007/s00220-017-3075-7} {\bibfield
  {journal} {\bibinfo  {journal} {Mathematics}\ }\textbf {\bibinfo {volume}
  {357}},\ \bibinfo {pages} {1253} (\bibinfo {year}
  {2018}{\natexlab{a}})}\BibitemShut {NoStop}%
\bibitem [{\citenamefont {Yunger~Halpern}\ \emph {et~al.}(2019)\citenamefont
  {Yunger~Halpern}, \citenamefont {Bartolotta},\ and\ \citenamefont
  {Pollack}}]{yunger2019entropic}%
  \BibitemOpen
  \bibfield  {author} {\bibinfo {author} {\bibfnamefont {N.}~\bibnamefont
  {Yunger~Halpern}}, \bibinfo {author} {\bibfnamefont {A.}~\bibnamefont
  {Bartolotta}},\ and\ \bibinfo {author} {\bibfnamefont {J.}~\bibnamefont
  {Pollack}},\ }\bibfield  {title} {\bibinfo {title} {Entropic uncertainty
  relations for quantum information scrambling},\ }\href
  {https://doi.org/https://doi.org/10.1038/s42005-019-0179-8} {\bibfield
  {journal} {\bibinfo  {journal} {Communications Physics}\ }\textbf {\bibinfo
  {volume} {2}},\ \bibinfo {pages} {1} (\bibinfo {year} {2019})}\BibitemShut
  {NoStop}%
\bibitem [{\citenamefont {Heinosaari}\ \emph {et~al.}(2016)\citenamefont
  {Heinosaari}, \citenamefont {Miyadera},\ and\ \citenamefont
  {Ziman}}]{Heinosaari_2016}%
  \BibitemOpen
  \bibfield  {author} {\bibinfo {author} {\bibfnamefont {T.}~\bibnamefont
  {Heinosaari}}, \bibinfo {author} {\bibfnamefont {T.}~\bibnamefont
  {Miyadera}},\ and\ \bibinfo {author} {\bibfnamefont {M.}~\bibnamefont
  {Ziman}},\ }\bibfield  {title} {\bibinfo {title} {An invitation to quantum
  incompatibility},\ }\href {https://doi.org/10.1088/1751-8113/49/12/123001}
  {\bibfield  {journal} {\bibinfo  {journal} {Journal of Physics A:
  Mathematical and Theoretical}\ }\textbf {\bibinfo {volume} {49}},\ \bibinfo
  {pages} {123001} (\bibinfo {year} {2016})}\BibitemShut {NoStop}%
\bibitem [{\citenamefont {G\"uhne}\ \emph {et~al.}(2021)\citenamefont
  {G\"uhne}, \citenamefont {Haapasalo}, \citenamefont {Kraft}, \citenamefont
  {Pellonp\"a\"a},\ and\ \citenamefont {Uola}}]{Guhne2021}%
  \BibitemOpen
  \bibfield  {author} {\bibinfo {author} {\bibfnamefont {O.}~\bibnamefont
  {G\"uhne}}, \bibinfo {author} {\bibfnamefont {E.}~\bibnamefont {Haapasalo}},
  \bibinfo {author} {\bibfnamefont {T.}~\bibnamefont {Kraft}}, \bibinfo
  {author} {\bibfnamefont {J.-P.}\ \bibnamefont {Pellonp\"a\"a}},\ and\
  \bibinfo {author} {\bibfnamefont {R.}~\bibnamefont {Uola}},\ }\bibfield
  {title} {\bibinfo {title} {{Incompatible measurements in quantum information
  science}},\ }\href {https://arxiv.org/abs/2112.06784} {\bibfield  {journal}
  {\bibinfo  {journal} {arXiv preprint:2112.06784}\ } (\bibinfo {year}
  {2021})}\BibitemShut {NoStop}%
\bibitem [{\citenamefont {Busch}\ \emph {et~al.}(2007)\citenamefont {Busch},
  \citenamefont {Heinonen},\ and\ \citenamefont {Lahti}}]{BUSCH2007155}%
  \BibitemOpen
  \bibfield  {author} {\bibinfo {author} {\bibfnamefont {P.}~\bibnamefont
  {Busch}}, \bibinfo {author} {\bibfnamefont {T.}~\bibnamefont {Heinonen}},\
  and\ \bibinfo {author} {\bibfnamefont {P.}~\bibnamefont {Lahti}},\ }\bibfield
   {title} {\bibinfo {title} {Heisenberg's uncertainty principle},\ }\href
  {https://doi.org/https://doi.org/10.1016/j.physrep.2007.05.006} {\bibfield
  {journal} {\bibinfo  {journal} {Physics Reports}\ }\textbf {\bibinfo {volume}
  {452}},\ \bibinfo {pages} {155} (\bibinfo {year} {2007})}\BibitemShut
  {NoStop}%
\bibitem [{\citenamefont {Clerk}\ \emph {et~al.}(2010)\citenamefont {Clerk},
  \citenamefont {Devoret}, \citenamefont {Girvin}, \citenamefont {Marquardt},\
  and\ \citenamefont {Schoelkopf}}]{RevModPhys.82.1155}%
  \BibitemOpen
  \bibfield  {author} {\bibinfo {author} {\bibfnamefont {A.~A.}\ \bibnamefont
  {Clerk}}, \bibinfo {author} {\bibfnamefont {M.~H.}\ \bibnamefont {Devoret}},
  \bibinfo {author} {\bibfnamefont {S.~M.}\ \bibnamefont {Girvin}}, \bibinfo
  {author} {\bibfnamefont {F.}~\bibnamefont {Marquardt}},\ and\ \bibinfo
  {author} {\bibfnamefont {R.~J.}\ \bibnamefont {Schoelkopf}},\ }\bibfield
  {title} {\bibinfo {title} {Introduction to quantum noise, measurement, and
  amplification},\ }\href {https://doi.org/10.1103/RevModPhys.82.1155}
  {\bibfield  {journal} {\bibinfo  {journal} {Rev. Mod. Phys.}\ }\textbf
  {\bibinfo {volume} {82}},\ \bibinfo {pages} {1155} (\bibinfo {year}
  {2010})}\BibitemShut {NoStop}%
\bibitem [{\citenamefont {Giovannetti}\ \emph {et~al.}(2011)\citenamefont
  {Giovannetti}, \citenamefont {Lloyd},\ and\ \citenamefont
  {Maccone}}]{Giovannetti2011nphoton}%
  \BibitemOpen
  \bibfield  {author} {\bibinfo {author} {\bibfnamefont {V.}~\bibnamefont
  {Giovannetti}}, \bibinfo {author} {\bibfnamefont {S.}~\bibnamefont {Lloyd}},\
  and\ \bibinfo {author} {\bibfnamefont {L.}~\bibnamefont {Maccone}},\
  }\bibfield  {title} {\bibinfo {title} {Advances in quantum metrology},\
  }\href {https://doi.org/10.1038/nphoton.2011.35} {\bibfield  {journal}
  {\bibinfo  {journal} {Nat. Photon.}\ }\textbf {\bibinfo {volume} {5}},\
  \bibinfo {pages} {222} (\bibinfo {year} {2011})}\BibitemShut {NoStop}%
\bibitem [{\citenamefont {Liang}\ \emph {et~al.}(2011)\citenamefont {Liang},
  \citenamefont {Spekkens},\ and\ \citenamefont {Wiseman}}]{LIANG20111}%
  \BibitemOpen
  \bibfield  {author} {\bibinfo {author} {\bibfnamefont {Y.-C.}\ \bibnamefont
  {Liang}}, \bibinfo {author} {\bibfnamefont {R.~W.}\ \bibnamefont
  {Spekkens}},\ and\ \bibinfo {author} {\bibfnamefont {H.~M.}\ \bibnamefont
  {Wiseman}},\ }\bibfield  {title} {\bibinfo {title} {Specker’s parable of
  the overprotective seer: A road to contextuality, nonlocality and
  complementarity},\ }\href
  {https://doi.org/https://doi.org/10.1016/j.physrep.2011.05.001} {\bibfield
  {journal} {\bibinfo  {journal} {Physics Reports}\ }\textbf {\bibinfo {volume}
  {506}},\ \bibinfo {pages} {1} (\bibinfo {year} {2011})}\BibitemShut {NoStop}%
\bibitem [{\citenamefont {Modi}\ \emph {et~al.}(2012)\citenamefont {Modi},
  \citenamefont {Brodutch}, \citenamefont {Cable}, \citenamefont {Paterek},\
  and\ \citenamefont {Vedral}}]{RevModPhys.84.1655}%
  \BibitemOpen
  \bibfield  {author} {\bibinfo {author} {\bibfnamefont {K.}~\bibnamefont
  {Modi}}, \bibinfo {author} {\bibfnamefont {A.}~\bibnamefont {Brodutch}},
  \bibinfo {author} {\bibfnamefont {H.}~\bibnamefont {Cable}}, \bibinfo
  {author} {\bibfnamefont {T.}~\bibnamefont {Paterek}},\ and\ \bibinfo {author}
  {\bibfnamefont {V.}~\bibnamefont {Vedral}},\ }\bibfield  {title} {\bibinfo
  {title} {The classical-quantum boundary for correlations: Discord and related
  measures},\ }\href {https://doi.org/10.1103/RevModPhys.84.1655} {\bibfield
  {journal} {\bibinfo  {journal} {Rev. Mod. Phys.}\ }\textbf {\bibinfo {volume}
  {84}},\ \bibinfo {pages} {1655} (\bibinfo {year} {2012})}\BibitemShut
  {NoStop}%
\bibitem [{\citenamefont {Chitambar}\ and\ \citenamefont
  {Gour}(2019)}]{RevModPhys.91.025001}%
  \BibitemOpen
  \bibfield  {author} {\bibinfo {author} {\bibfnamefont {E.}~\bibnamefont
  {Chitambar}}\ and\ \bibinfo {author} {\bibfnamefont {G.}~\bibnamefont
  {Gour}},\ }\bibfield  {title} {\bibinfo {title} {Quantum resource theories},\
  }\href {https://doi.org/10.1103/RevModPhys.91.025001} {\bibfield  {journal}
  {\bibinfo  {journal} {Rev. Mod. Phys.}\ }\textbf {\bibinfo {volume} {91}},\
  \bibinfo {pages} {025001} (\bibinfo {year} {2019})}\BibitemShut {NoStop}%
\bibitem [{\citenamefont {Portmann}\ and\ \citenamefont
  {Renner}(2022)}]{RevModPhys.94.025008}%
  \BibitemOpen
  \bibfield  {author} {\bibinfo {author} {\bibfnamefont {C.}~\bibnamefont
  {Portmann}}\ and\ \bibinfo {author} {\bibfnamefont {R.}~\bibnamefont
  {Renner}},\ }\bibfield  {title} {\bibinfo {title} {Security in quantum
  cryptography},\ }\href {https://doi.org/10.1103/RevModPhys.94.025008}
  {\bibfield  {journal} {\bibinfo  {journal} {Rev. Mod. Phys.}\ }\textbf
  {\bibinfo {volume} {94}},\ \bibinfo {pages} {025008} (\bibinfo {year}
  {2022})}\BibitemShut {NoStop}%
\bibitem [{\citenamefont {Robertson}(1934)}]{PhysRev.46.794}%
  \BibitemOpen
  \bibfield  {author} {\bibinfo {author} {\bibfnamefont {H.~P.}\ \bibnamefont
  {Robertson}},\ }\bibfield  {title} {\bibinfo {title} {An indeterminacy
  relation for several observables and its classical interpretation},\ }\href
  {https://doi.org/10.1103/PhysRev.46.794} {\bibfield  {journal} {\bibinfo
  {journal} {Phys. Rev.}\ }\textbf {\bibinfo {volume} {46}},\ \bibinfo {pages}
  {794} (\bibinfo {year} {1934})}\BibitemShut {NoStop}%
\bibitem [{\citenamefont {Wehner}\ and\ \citenamefont
  {Winter}(2008)}]{Winter2008}%
  \BibitemOpen
  \bibfield  {author} {\bibinfo {author} {\bibfnamefont {S.}~\bibnamefont
  {Wehner}}\ and\ \bibinfo {author} {\bibfnamefont {A.}~\bibnamefont
  {Winter}},\ }\bibfield  {title} {\bibinfo {title} {Higher entropic
  uncertainty relations for anti-commuting observables},\ }\href
  {https://doi.org/10.1063/1.2943685} {\bibfield  {journal} {\bibinfo
  {journal} {Journal of Mathematical Physics}\ }\textbf {\bibinfo {volume}
  {49}},\ \bibinfo {pages} {062105} (\bibinfo {year} {2008})}\BibitemShut
  {NoStop}%
\bibitem [{\citenamefont {Huang}(2012)}]{PhysRevA.86.024101}%
  \BibitemOpen
  \bibfield  {author} {\bibinfo {author} {\bibfnamefont {Y.}~\bibnamefont
  {Huang}},\ }\bibfield  {title} {\bibinfo {title} {Variance-based uncertainty
  relations},\ }\href {https://doi.org/10.1103/PhysRevA.86.024101} {\bibfield
  {journal} {\bibinfo  {journal} {Phys. Rev. A}\ }\textbf {\bibinfo {volume}
  {86}},\ \bibinfo {pages} {024101} (\bibinfo {year} {2012})}\BibitemShut
  {NoStop}%
\bibitem [{\citenamefont {Abbott}\ \emph {et~al.}(2016)\citenamefont {Abbott},
  \citenamefont {Alzieu}, \citenamefont {Hall},\ and\ \citenamefont
  {Branciard}}]{math4010008}%
  \BibitemOpen
  \bibfield  {author} {\bibinfo {author} {\bibfnamefont {A.~A.}\ \bibnamefont
  {Abbott}}, \bibinfo {author} {\bibfnamefont {P.-L.}\ \bibnamefont {Alzieu}},
  \bibinfo {author} {\bibfnamefont {M.~J.~W.}\ \bibnamefont {Hall}},\ and\
  \bibinfo {author} {\bibfnamefont {C.}~\bibnamefont {Branciard}},\ }\bibfield
  {title} {\bibinfo {title} {Tight state-independent uncertainty relations for
  qubits},\ }\href {https://doi.org/10.3390/math4010008} {\bibfield  {journal}
  {\bibinfo  {journal} {Mathematics}\ }\textbf {\bibinfo {volume} {4}},\
  \bibinfo {pages} {8} (\bibinfo {year} {2016})}\BibitemShut {NoStop}%
\bibitem [{\citenamefont {Maccone}\ and\ \citenamefont
  {Pati}(2014)}]{Maccone2014PRL}%
  \BibitemOpen
  \bibfield  {author} {\bibinfo {author} {\bibfnamefont {L.}~\bibnamefont
  {Maccone}}\ and\ \bibinfo {author} {\bibfnamefont {A.~K.}\ \bibnamefont
  {Pati}},\ }\bibfield  {title} {\bibinfo {title} {Stronger uncertainty
  relations for all incompatible observables},\ }\href
  {https://doi.org/10.1103/PhysRevLett.113.260401} {\bibfield  {journal}
  {\bibinfo  {journal} {Phys. Rev. Lett.}\ }\textbf {\bibinfo {volume} {113}},\
  \bibinfo {pages} {260401} (\bibinfo {year} {2014})}\BibitemShut {NoStop}%
\bibitem [{\citenamefont {Kechrimparis}\ and\ \citenamefont
  {Weigert}(2014)}]{Kechrimparis2014PRA}%
  \BibitemOpen
  \bibfield  {author} {\bibinfo {author} {\bibfnamefont {S.}~\bibnamefont
  {Kechrimparis}}\ and\ \bibinfo {author} {\bibfnamefont {S.}~\bibnamefont
  {Weigert}},\ }\bibfield  {title} {\bibinfo {title} {Heisenberg uncertainty
  relation for three canonical observables},\ }\href
  {https://doi.org/10.1103/PhysRevA.90.062118} {\bibfield  {journal} {\bibinfo
  {journal} {Phys. Rev. A}\ }\textbf {\bibinfo {volume} {90}},\ \bibinfo
  {pages} {062118} (\bibinfo {year} {2014})}\BibitemShut {NoStop}%
\bibitem [{\citenamefont {Ma}\ \emph {et~al.}(2017)\citenamefont {Ma},
  \citenamefont {Chen}, \citenamefont {Liu}, \citenamefont {Wang},
  \citenamefont {Ye}, \citenamefont {Kong}, \citenamefont {Shi}, \citenamefont
  {Fei},\ and\ \citenamefont {Du}}]{Ma2017PRL}%
  \BibitemOpen
  \bibfield  {author} {\bibinfo {author} {\bibfnamefont {W.}~\bibnamefont
  {Ma}}, \bibinfo {author} {\bibfnamefont {B.}~\bibnamefont {Chen}}, \bibinfo
  {author} {\bibfnamefont {Y.}~\bibnamefont {Liu}}, \bibinfo {author}
  {\bibfnamefont {M.}~\bibnamefont {Wang}}, \bibinfo {author} {\bibfnamefont
  {X.}~\bibnamefont {Ye}}, \bibinfo {author} {\bibfnamefont {F.}~\bibnamefont
  {Kong}}, \bibinfo {author} {\bibfnamefont {F.}~\bibnamefont {Shi}}, \bibinfo
  {author} {\bibfnamefont {S.-M.}\ \bibnamefont {Fei}},\ and\ \bibinfo {author}
  {\bibfnamefont {J.}~\bibnamefont {Du}},\ }\bibfield  {title} {\bibinfo
  {title} {Experimental demonstration of uncertainty relations for the triple
  components of angular momentum},\ }\href
  {https://doi.org/10.1103/PhysRevLett.118.180402} {\bibfield  {journal}
  {\bibinfo  {journal} {Phys. Rev. Lett.}\ }\textbf {\bibinfo {volume} {118}},\
  \bibinfo {pages} {180402} (\bibinfo {year} {2017})}\BibitemShut {NoStop}%
\bibitem [{\citenamefont {Fan}\ \emph {et~al.}(2018)\citenamefont {Fan},
  \citenamefont {Wang}, \citenamefont {Xiao},\ and\ \citenamefont
  {Xue}}]{Fan2018PRA}%
  \BibitemOpen
  \bibfield  {author} {\bibinfo {author} {\bibfnamefont {B.}~\bibnamefont
  {Fan}}, \bibinfo {author} {\bibfnamefont {K.}~\bibnamefont {Wang}}, \bibinfo
  {author} {\bibfnamefont {L.}~\bibnamefont {Xiao}},\ and\ \bibinfo {author}
  {\bibfnamefont {P.}~\bibnamefont {Xue}},\ }\bibfield  {title} {\bibinfo
  {title} {Experimental test of a stronger multiobservable uncertainty
  relation},\ }\href {https://doi.org/10.1103/PhysRevA.98.032118} {\bibfield
  {journal} {\bibinfo  {journal} {Phys. Rev. A}\ }\textbf {\bibinfo {volume}
  {98}},\ \bibinfo {pages} {032118} (\bibinfo {year} {2018})}\BibitemShut
  {NoStop}%
\bibitem [{\citenamefont {Yu}\ and\ \citenamefont {Oh}(2013)}]{Yu2013arXiv}%
  \BibitemOpen
  \bibfield  {author} {\bibinfo {author} {\bibfnamefont {S.}~\bibnamefont
  {Yu}}\ and\ \bibinfo {author} {\bibfnamefont {C.}~\bibnamefont {Oh}},\
  }\bibfield  {title} {\bibinfo {title} {Quantum contextuality and joint
  measurement of three observables of a qubit},\ }\href
  {https://arxiv.org/abs/1312.6470} {\bibfield  {journal} {\bibinfo  {journal}
  {arXiv preprint arXiv:1312.6470}\ } (\bibinfo {year} {2013})}\BibitemShut
  {NoStop}%
\bibitem [{\citenamefont {Schwonnek}\ \emph {et~al.}(2016)\citenamefont
  {Schwonnek}, \citenamefont {Reeb},\ and\ \citenamefont
  {Werner}}]{Schwonnek2016math4020038}%
  \BibitemOpen
  \bibfield  {author} {\bibinfo {author} {\bibfnamefont {R.}~\bibnamefont
  {Schwonnek}}, \bibinfo {author} {\bibfnamefont {D.}~\bibnamefont {Reeb}},\
  and\ \bibinfo {author} {\bibfnamefont {R.~F.}\ \bibnamefont {Werner}},\
  }\bibfield  {title} {\bibinfo {title} {Measurement uncertainty for finite
  quantum observables},\ }\href {https://doi.org/10.3390/math4020038}
  {\bibfield  {journal} {\bibinfo  {journal} {Mathematics}\ }\textbf {\bibinfo
  {volume} {4}},\ \bibinfo {pages} {38} (\bibinfo {year} {2016})}\BibitemShut
  {NoStop}%
\bibitem [{\citenamefont {Zhao}\ \emph {et~al.}(2017)\citenamefont {Zhao},
  \citenamefont {Kurzy\ifmmode~\acute{n}\else \'{n}\fi{}ski}, \citenamefont
  {Xiang}, \citenamefont {Li},\ and\ \citenamefont {Guo}}]{Zhao2017PRA}%
  \BibitemOpen
  \bibfield  {author} {\bibinfo {author} {\bibfnamefont {Y.-Y.}\ \bibnamefont
  {Zhao}}, \bibinfo {author} {\bibfnamefont {P.}~\bibnamefont
  {Kurzy\ifmmode~\acute{n}\else \'{n}\fi{}ski}}, \bibinfo {author}
  {\bibfnamefont {G.-Y.}\ \bibnamefont {Xiang}}, \bibinfo {author}
  {\bibfnamefont {C.-F.}\ \bibnamefont {Li}},\ and\ \bibinfo {author}
  {\bibfnamefont {G.-C.}\ \bibnamefont {Guo}},\ }\bibfield  {title} {\bibinfo
  {title} {Heisenberg's error-disturbance relations: A joint measurement-based
  experimental test},\ }\href {https://doi.org/10.1103/PhysRevA.95.040101}
  {\bibfield  {journal} {\bibinfo  {journal} {Phys. Rev. A}\ }\textbf {\bibinfo
  {volume} {95}},\ \bibinfo {pages} {040101} (\bibinfo {year}
  {2017})}\BibitemShut {NoStop}%
\bibitem [{\citenamefont {Qin}\ \emph {et~al.}(2019)\citenamefont {Qin},
  \citenamefont {Zhang}, \citenamefont {Jost}, \citenamefont {Sun},
  \citenamefont {Li-Jost},\ and\ \citenamefont {Fei}}]{Qin2019PRA}%
  \BibitemOpen
  \bibfield  {author} {\bibinfo {author} {\bibfnamefont {H.-H.}\ \bibnamefont
  {Qin}}, \bibinfo {author} {\bibfnamefont {T.-G.}\ \bibnamefont {Zhang}},
  \bibinfo {author} {\bibfnamefont {L.}~\bibnamefont {Jost}}, \bibinfo {author}
  {\bibfnamefont {C.-P.}\ \bibnamefont {Sun}}, \bibinfo {author} {\bibfnamefont
  {X.}~\bibnamefont {Li-Jost}},\ and\ \bibinfo {author} {\bibfnamefont {S.-M.}\
  \bibnamefont {Fei}},\ }\bibfield  {title} {\bibinfo {title} {Uncertainties of
  genuinely incompatible triple measurements based on statistical distance},\
  }\href {https://doi.org/10.1103/PhysRevA.99.032107} {\bibfield  {journal}
  {\bibinfo  {journal} {Phys. Rev. A}\ }\textbf {\bibinfo {volume} {99}},\
  \bibinfo {pages} {032107} (\bibinfo {year} {2019})}\BibitemShut {NoStop}%
\bibitem [{\citenamefont {Qin}\ and\ \citenamefont {Fei}(2022)}]{EP2022}%
  \BibitemOpen
  \bibfield  {author} {\bibinfo {author} {\bibfnamefont {H.}~\bibnamefont
  {Qin}}\ and\ \bibinfo {author} {\bibfnamefont {S.-M.}\ \bibnamefont {Fei}},\
  }\bibfield  {title} {\bibinfo {title} {Optimizing incompatible triple quantum
  measurements},\ }\href {https://doi.org/10.1140/epjp/s13360-022-02826-0}
  {\bibfield  {journal} {\bibinfo  {journal} {The European Physical Journal
  Plus}\ }\textbf {\bibinfo {volume} {137}},\ \bibinfo {pages} {635} (\bibinfo
  {year} {2022})}\BibitemShut {NoStop}%
\bibitem [{\citenamefont {Lund}\ and\ \citenamefont
  {Wiseman}(2010)}]{Lund2010NJP}%
  \BibitemOpen
  \bibfield  {author} {\bibinfo {author} {\bibfnamefont {A.~P.}\ \bibnamefont
  {Lund}}\ and\ \bibinfo {author} {\bibfnamefont {H.~M.}\ \bibnamefont
  {Wiseman}},\ }\bibfield  {title} {\bibinfo {title} {Measuring
  measurement{\textendash}disturbance relationships with weak values},\ }\href
  {https://doi.org/10.1088/1367-2630/12/9/093011} {\bibfield  {journal}
  {\bibinfo  {journal} {New Journal of Physics}\ }\textbf {\bibinfo {volume}
  {12}},\ \bibinfo {pages} {093011} (\bibinfo {year} {2010})}\BibitemShut
  {NoStop}%
\bibitem [{\citenamefont {Di~Lorenzo}(2013)}]{PhysRevLett.110.010404}%
  \BibitemOpen
  \bibfield  {author} {\bibinfo {author} {\bibfnamefont {A.}~\bibnamefont
  {Di~Lorenzo}},\ }\bibfield  {title} {\bibinfo {title} {Sequential measurement
  of conjugate variables as an alternative quantum state tomography},\ }\href
  {https://doi.org/10.1103/PhysRevLett.110.010404} {\bibfield  {journal}
  {\bibinfo  {journal} {Phys. Rev. Lett.}\ }\textbf {\bibinfo {volume} {110}},\
  \bibinfo {pages} {010404} (\bibinfo {year} {2013})}\BibitemShut {NoStop}%
\bibitem [{\citenamefont {Erhart}\ \emph {et~al.}(2012)\citenamefont {Erhart},
  \citenamefont {Sponar}, \citenamefont {Sulyok}, \citenamefont {Badurek},
  \citenamefont {Ozawa},\ and\ \citenamefont {Hasegawa}}]{Erhart2012Nphysics}%
  \BibitemOpen
  \bibfield  {author} {\bibinfo {author} {\bibfnamefont {J.}~\bibnamefont
  {Erhart}}, \bibinfo {author} {\bibfnamefont {S.}~\bibnamefont {Sponar}},
  \bibinfo {author} {\bibfnamefont {G.}~\bibnamefont {Sulyok}}, \bibinfo
  {author} {\bibfnamefont {G.}~\bibnamefont {Badurek}}, \bibinfo {author}
  {\bibfnamefont {M.}~\bibnamefont {Ozawa}},\ and\ \bibinfo {author}
  {\bibfnamefont {Y.}~\bibnamefont {Hasegawa}},\ }\bibfield  {title} {\bibinfo
  {title} {Experimental demonstration of a universally valid error--disturbance
  uncertainty relation in spin measurements},\ }\href
  {https://doi.org/10.1038/nphys2194} {\bibfield  {journal} {\bibinfo
  {journal} {Nature Physics}\ }\textbf {\bibinfo {volume} {8}},\ \bibinfo
  {pages} {185} (\bibinfo {year} {2012})}\BibitemShut {NoStop}%
\bibitem [{\citenamefont {Rozema}\ \emph {et~al.}(2012)\citenamefont {Rozema},
  \citenamefont {Darabi}, \citenamefont {Mahler}, \citenamefont {Hayat},
  \citenamefont {Soudagar},\ and\ \citenamefont {Steinberg}}]{Rozema2012PRL}%
  \BibitemOpen
  \bibfield  {author} {\bibinfo {author} {\bibfnamefont {L.~A.}\ \bibnamefont
  {Rozema}}, \bibinfo {author} {\bibfnamefont {A.}~\bibnamefont {Darabi}},
  \bibinfo {author} {\bibfnamefont {D.~H.}\ \bibnamefont {Mahler}}, \bibinfo
  {author} {\bibfnamefont {A.}~\bibnamefont {Hayat}}, \bibinfo {author}
  {\bibfnamefont {Y.}~\bibnamefont {Soudagar}},\ and\ \bibinfo {author}
  {\bibfnamefont {A.~M.}\ \bibnamefont {Steinberg}},\ }\bibfield  {title}
  {\bibinfo {title} {Violation of {H}eisenberg's measurement-disturbance
  relationship by weak measurements},\ }\href
  {https://doi.org/10.1103/PhysRevLett.109.100404} {\bibfield  {journal}
  {\bibinfo  {journal} {Phys. Rev. Lett.}\ }\textbf {\bibinfo {volume} {109}},\
  \bibinfo {pages} {100404} (\bibinfo {year} {2012})}\BibitemShut {NoStop}%
\bibitem [{\citenamefont {Sulyok}\ \emph {et~al.}(2013)\citenamefont {Sulyok},
  \citenamefont {Sponar}, \citenamefont {Erhart}, \citenamefont {Badurek},
  \citenamefont {Ozawa},\ and\ \citenamefont {Hasegawa}}]{Sulyok2013PRA}%
  \BibitemOpen
  \bibfield  {author} {\bibinfo {author} {\bibfnamefont {G.}~\bibnamefont
  {Sulyok}}, \bibinfo {author} {\bibfnamefont {S.}~\bibnamefont {Sponar}},
  \bibinfo {author} {\bibfnamefont {J.}~\bibnamefont {Erhart}}, \bibinfo
  {author} {\bibfnamefont {G.}~\bibnamefont {Badurek}}, \bibinfo {author}
  {\bibfnamefont {M.}~\bibnamefont {Ozawa}},\ and\ \bibinfo {author}
  {\bibfnamefont {Y.}~\bibnamefont {Hasegawa}},\ }\bibfield  {title} {\bibinfo
  {title} {Violation of {H}eisenberg's error-disturbance uncertainty relation
  in neutron-spin measurements},\ }\href
  {https://doi.org/10.1103/PhysRevA.88.022110} {\bibfield  {journal} {\bibinfo
  {journal} {Phys. Rev. A}\ }\textbf {\bibinfo {volume} {88}},\ \bibinfo
  {pages} {022110} (\bibinfo {year} {2013})}\BibitemShut {NoStop}%
\bibitem [{\citenamefont {Baek}\ \emph {et~al.}(2013)\citenamefont {Baek},
  \citenamefont {Kaneda}, \citenamefont {Ozawa},\ and\ \citenamefont
  {Edamatsu}}]{Baek2013SP}%
  \BibitemOpen
  \bibfield  {author} {\bibinfo {author} {\bibfnamefont {S.-Y.}\ \bibnamefont
  {Baek}}, \bibinfo {author} {\bibfnamefont {F.}~\bibnamefont {Kaneda}},
  \bibinfo {author} {\bibfnamefont {M.}~\bibnamefont {Ozawa}},\ and\ \bibinfo
  {author} {\bibfnamefont {K.}~\bibnamefont {Edamatsu}},\ }\bibfield  {title}
  {\bibinfo {title} {Experimental violation and reformulation of the
  {H}eisenberg's error-disturbance uncertainty relation},\ }\href
  {https://doi.org/10.1038/srep02221} {\bibfield  {journal} {\bibinfo
  {journal} {Scientific reports}\ }\textbf {\bibinfo {volume} {3}},\ \bibinfo
  {pages} {2221} (\bibinfo {year} {2013})}\BibitemShut {NoStop}%
\bibitem [{\citenamefont {Weston}\ \emph {et~al.}(2013)\citenamefont {Weston},
  \citenamefont {Hall}, \citenamefont {Palsson}, \citenamefont {Wiseman},\ and\
  \citenamefont {Pryde}}]{PhysRevLett.110.220402}%
  \BibitemOpen
  \bibfield  {author} {\bibinfo {author} {\bibfnamefont {M.~M.}\ \bibnamefont
  {Weston}}, \bibinfo {author} {\bibfnamefont {M.~J.~W.}\ \bibnamefont {Hall}},
  \bibinfo {author} {\bibfnamefont {M.~S.}\ \bibnamefont {Palsson}}, \bibinfo
  {author} {\bibfnamefont {H.~M.}\ \bibnamefont {Wiseman}},\ and\ \bibinfo
  {author} {\bibfnamefont {G.~J.}\ \bibnamefont {Pryde}},\ }\bibfield  {title}
  {\bibinfo {title} {Experimental test of universal complementarity
  relations},\ }\href {https://doi.org/10.1103/PhysRevLett.110.220402}
  {\bibfield  {journal} {\bibinfo  {journal} {Phys. Rev. Lett.}\ }\textbf
  {\bibinfo {volume} {110}},\ \bibinfo {pages} {220402} (\bibinfo {year}
  {2013})}\BibitemShut {NoStop}%
\bibitem [{\citenamefont {Ringbauer}\ \emph {et~al.}(2014)\citenamefont
  {Ringbauer}, \citenamefont {Biggerstaff}, \citenamefont {Broome},
  \citenamefont {Fedrizzi}, \citenamefont {Branciard},\ and\ \citenamefont
  {White}}]{Ringbauer2014PRL}%
  \BibitemOpen
  \bibfield  {author} {\bibinfo {author} {\bibfnamefont {M.}~\bibnamefont
  {Ringbauer}}, \bibinfo {author} {\bibfnamefont {D.~N.}\ \bibnamefont
  {Biggerstaff}}, \bibinfo {author} {\bibfnamefont {M.~A.}\ \bibnamefont
  {Broome}}, \bibinfo {author} {\bibfnamefont {A.}~\bibnamefont {Fedrizzi}},
  \bibinfo {author} {\bibfnamefont {C.}~\bibnamefont {Branciard}},\ and\
  \bibinfo {author} {\bibfnamefont {A.~G.}\ \bibnamefont {White}},\ }\bibfield
  {title} {\bibinfo {title} {Experimental joint quantum measurements with
  minimum uncertainty},\ }\href
  {https://doi.org/10.1103/PhysRevLett.112.020401} {\bibfield  {journal}
  {\bibinfo  {journal} {Phys. Rev. Lett.}\ }\textbf {\bibinfo {volume} {112}},\
  \bibinfo {pages} {020401} (\bibinfo {year} {2014})}\BibitemShut {NoStop}%
\bibitem [{\citenamefont {Kaneda}\ \emph {et~al.}(2014)\citenamefont {Kaneda},
  \citenamefont {Baek}, \citenamefont {Ozawa},\ and\ \citenamefont
  {Edamatsu}}]{Kaneda2014PRL}%
  \BibitemOpen
  \bibfield  {author} {\bibinfo {author} {\bibfnamefont {F.}~\bibnamefont
  {Kaneda}}, \bibinfo {author} {\bibfnamefont {S.-Y.}\ \bibnamefont {Baek}},
  \bibinfo {author} {\bibfnamefont {M.}~\bibnamefont {Ozawa}},\ and\ \bibinfo
  {author} {\bibfnamefont {K.}~\bibnamefont {Edamatsu}},\ }\bibfield  {title}
  {\bibinfo {title} {Experimental test of error-disturbance uncertainty
  relations by weak measurement},\ }\href
  {https://doi.org/10.1103/PhysRevLett.112.020402} {\bibfield  {journal}
  {\bibinfo  {journal} {Phys. Rev. Lett.}\ }\textbf {\bibinfo {volume} {112}},\
  \bibinfo {pages} {020402} (\bibinfo {year} {2014})}\BibitemShut {NoStop}%
\bibitem [{\citenamefont {Ma}\ \emph {et~al.}(2016)\citenamefont {Ma},
  \citenamefont {Ma}, \citenamefont {Wang}, \citenamefont {Chen}, \citenamefont
  {Liu}, \citenamefont {Kong}, \citenamefont {Li}, \citenamefont {Peng},
  \citenamefont {Shi}, \citenamefont {Shi}, \citenamefont {Fei},\ and\
  \citenamefont {Du}}]{Ma2016PRL}%
  \BibitemOpen
  \bibfield  {author} {\bibinfo {author} {\bibfnamefont {W.}~\bibnamefont
  {Ma}}, \bibinfo {author} {\bibfnamefont {Z.}~\bibnamefont {Ma}}, \bibinfo
  {author} {\bibfnamefont {H.}~\bibnamefont {Wang}}, \bibinfo {author}
  {\bibfnamefont {Z.}~\bibnamefont {Chen}}, \bibinfo {author} {\bibfnamefont
  {Y.}~\bibnamefont {Liu}}, \bibinfo {author} {\bibfnamefont {F.}~\bibnamefont
  {Kong}}, \bibinfo {author} {\bibfnamefont {Z.}~\bibnamefont {Li}}, \bibinfo
  {author} {\bibfnamefont {X.}~\bibnamefont {Peng}}, \bibinfo {author}
  {\bibfnamefont {M.}~\bibnamefont {Shi}}, \bibinfo {author} {\bibfnamefont
  {F.}~\bibnamefont {Shi}}, \bibinfo {author} {\bibfnamefont {S.-M.}\
  \bibnamefont {Fei}},\ and\ \bibinfo {author} {\bibfnamefont {J.}~\bibnamefont
  {Du}},\ }\bibfield  {title} {\bibinfo {title} {Experimental test of
  {H}eisenberg's measurement uncertainty relation based on statistical
  distances},\ }\href {https://doi.org/10.1103/PhysRevLett.116.160405}
  {\bibfield  {journal} {\bibinfo  {journal} {Phys. Rev. Lett.}\ }\textbf
  {\bibinfo {volume} {116}},\ \bibinfo {pages} {160405} (\bibinfo {year}
  {2016})}\BibitemShut {NoStop}%
\bibitem [{\citenamefont {Zhou}\ \emph {et~al.}(2016)\citenamefont {Zhou},
  \citenamefont {Yan}, \citenamefont {Gong}, \citenamefont {Ma}, \citenamefont
  {He}, \citenamefont {Xiong}, \citenamefont {Chen}, \citenamefont {Yang},
  \citenamefont {Feng},\ and\ \citenamefont {Vedral}}]{Zhou2016SA}%
  \BibitemOpen
  \bibfield  {author} {\bibinfo {author} {\bibfnamefont {F.}~\bibnamefont
  {Zhou}}, \bibinfo {author} {\bibfnamefont {L.}~\bibnamefont {Yan}}, \bibinfo
  {author} {\bibfnamefont {S.}~\bibnamefont {Gong}}, \bibinfo {author}
  {\bibfnamefont {Z.}~\bibnamefont {Ma}}, \bibinfo {author} {\bibfnamefont
  {J.}~\bibnamefont {He}}, \bibinfo {author} {\bibfnamefont {T.}~\bibnamefont
  {Xiong}}, \bibinfo {author} {\bibfnamefont {L.}~\bibnamefont {Chen}},
  \bibinfo {author} {\bibfnamefont {W.}~\bibnamefont {Yang}}, \bibinfo {author}
  {\bibfnamefont {M.}~\bibnamefont {Feng}},\ and\ \bibinfo {author}
  {\bibfnamefont {V.}~\bibnamefont {Vedral}},\ }\bibfield  {title} {\bibinfo
  {title} {Verifying {H}eisenberg{\textquoteright}s error-disturbance relation
  using a single trapped ion},\ }\href {https://doi.org/10.1126/sciadv.1600578}
  {\bibfield  {journal} {\bibinfo  {journal} {Science Advances}\ }\textbf
  {\bibinfo {volume} {2}},\ \bibinfo {pages} {e1600578} (\bibinfo {year}
  {2016})}\BibitemShut {NoStop}%
\bibitem [{\citenamefont {Xiong}\ \emph {et~al.}(2017)\citenamefont {Xiong},
  \citenamefont {Yan}, \citenamefont {Ma}, \citenamefont {Zhou}, \citenamefont
  {Chen}, \citenamefont {Yang}, \citenamefont {Feng},\ and\ \citenamefont
  {Busch}}]{Xiong2017NJP}%
  \BibitemOpen
  \bibfield  {author} {\bibinfo {author} {\bibfnamefont {T.~P.}\ \bibnamefont
  {Xiong}}, \bibinfo {author} {\bibfnamefont {L.~L.}\ \bibnamefont {Yan}},
  \bibinfo {author} {\bibfnamefont {Z.~H.}\ \bibnamefont {Ma}}, \bibinfo
  {author} {\bibfnamefont {F.}~\bibnamefont {Zhou}}, \bibinfo {author}
  {\bibfnamefont {L.}~\bibnamefont {Chen}}, \bibinfo {author} {\bibfnamefont
  {W.~L.}\ \bibnamefont {Yang}}, \bibinfo {author} {\bibfnamefont
  {M.}~\bibnamefont {Feng}},\ and\ \bibinfo {author} {\bibfnamefont
  {P.}~\bibnamefont {Busch}},\ }\bibfield  {title} {\bibinfo {title} {Optimal
  joint measurements of complementary observables by a single trapped ion},\
  }\href {https://doi.org/10.1088/1367-2630/aa70a5} {\bibfield  {journal}
  {\bibinfo  {journal} {New Journal of Physics}\ }\textbf {\bibinfo {volume}
  {19}},\ \bibinfo {pages} {063032} (\bibinfo {year} {2017})}\BibitemShut
  {NoStop}%
\bibitem [{\citenamefont {Bullock}\ and\ \citenamefont
  {Busch}(2018{\natexlab{b}})}]{Bullock_2018}%
  \BibitemOpen
  \bibfield  {author} {\bibinfo {author} {\bibfnamefont {T.}~\bibnamefont
  {Bullock}}\ and\ \bibinfo {author} {\bibfnamefont {P.}~\bibnamefont
  {Busch}},\ }\bibfield  {title} {\bibinfo {title} {Measurement uncertainty
  relations: characterising optimal error bounds for qubits},\ }\href
  {https://doi.org/10.1088/1751-8121/aac729} {\bibfield  {journal} {\bibinfo
  {journal} {Journal of Physics A: Mathematical and Theoretical}\ }\textbf
  {\bibinfo {volume} {51}},\ \bibinfo {pages} {283001} (\bibinfo {year}
  {2018}{\natexlab{b}})}\BibitemShut {NoStop}%
\bibitem [{\citenamefont {Mao}\ \emph {et~al.}(2019)\citenamefont {Mao},
  \citenamefont {Ma}, \citenamefont {Jin}, \citenamefont {Sun}, \citenamefont
  {Fei}, \citenamefont {Zhang}, \citenamefont {Fan},\ and\ \citenamefont
  {Pan}}]{Mao2019PRL}%
  \BibitemOpen
  \bibfield  {author} {\bibinfo {author} {\bibfnamefont {Y.-L.}\ \bibnamefont
  {Mao}}, \bibinfo {author} {\bibfnamefont {Z.-H.}\ \bibnamefont {Ma}},
  \bibinfo {author} {\bibfnamefont {R.-B.}\ \bibnamefont {Jin}}, \bibinfo
  {author} {\bibfnamefont {Q.-C.}\ \bibnamefont {Sun}}, \bibinfo {author}
  {\bibfnamefont {S.-M.}\ \bibnamefont {Fei}}, \bibinfo {author} {\bibfnamefont
  {Q.}~\bibnamefont {Zhang}}, \bibinfo {author} {\bibfnamefont
  {J.}~\bibnamefont {Fan}},\ and\ \bibinfo {author} {\bibfnamefont {J.-W.}\
  \bibnamefont {Pan}},\ }\bibfield  {title} {\bibinfo {title}
  {Error-disturbance trade-off in sequential quantum measurements},\ }\href
  {https://doi.org/10.1103/PhysRevLett.122.090404} {\bibfield  {journal}
  {\bibinfo  {journal} {Phys. Rev. Lett.}\ }\textbf {\bibinfo {volume} {122}},\
  \bibinfo {pages} {090404} (\bibinfo {year} {2019})}\BibitemShut {NoStop}%
\bibitem [{\citenamefont {Monroe}\ \emph {et~al.}(2021)\citenamefont {Monroe},
  \citenamefont {Yunger~Halpern}, \citenamefont {Lee},\ and\ \citenamefont
  {Murch}}]{yungerExp}%
  \BibitemOpen
  \bibfield  {author} {\bibinfo {author} {\bibfnamefont {J.~T.}\ \bibnamefont
  {Monroe}}, \bibinfo {author} {\bibfnamefont {N.}~\bibnamefont
  {Yunger~Halpern}}, \bibinfo {author} {\bibfnamefont {T.}~\bibnamefont
  {Lee}},\ and\ \bibinfo {author} {\bibfnamefont {K.~W.}\ \bibnamefont
  {Murch}},\ }\bibfield  {title} {\bibinfo {title} {Weak measurement of a
  superconducting qubit reconciles incompatible operators},\ }\href
  {https://doi.org/10.1103/PhysRevLett.126.100403} {\bibfield  {journal}
  {\bibinfo  {journal} {Phys. Rev. Lett.}\ }\textbf {\bibinfo {volume} {126}},\
  \bibinfo {pages} {100403} (\bibinfo {year} {2021})}\BibitemShut {NoStop}%
\bibitem [{\citenamefont {Ali}\ \emph {et~al.}(2009)\citenamefont {Ali},
  \citenamefont {Carmeli}, \citenamefont {Heinosaari},\ and\ \citenamefont
  {Toigo}}]{Toigo2009}%
  \BibitemOpen
  \bibfield  {author} {\bibinfo {author} {\bibfnamefont {S.~T.}\ \bibnamefont
  {Ali}}, \bibinfo {author} {\bibfnamefont {C.}~\bibnamefont {Carmeli}},
  \bibinfo {author} {\bibfnamefont {T.}~\bibnamefont {Heinosaari}},\ and\
  \bibinfo {author} {\bibfnamefont {A.}~\bibnamefont {Toigo}},\ }\bibfield
  {title} {\bibinfo {title} {Commutative {POVM}s and fuzzy observables},\
  }\href {https://doi.org/10.1007/s10701-009-9292-y} {\bibfield  {journal}
  {\bibinfo  {journal} {Foundations of Physics}\ }\textbf {\bibinfo {volume}
  {39}},\ \bibinfo {pages} {593} (\bibinfo {year} {2009})}\BibitemShut
  {NoStop}%
\bibitem [{\citenamefont {Busch}(1986)}]{PhysRevD.33.2253}%
  \BibitemOpen
  \bibfield  {author} {\bibinfo {author} {\bibfnamefont {P.}~\bibnamefont
  {Busch}},\ }\bibfield  {title} {\bibinfo {title} {Unsharp reality and joint
  measurements for spin observables},\ }\href
  {https://doi.org/10.1103/PhysRevD.33.2253} {\bibfield  {journal} {\bibinfo
  {journal} {Phys. Rev. D}\ }\textbf {\bibinfo {volume} {33}},\ \bibinfo
  {pages} {2253} (\bibinfo {year} {1986})}\BibitemShut {NoStop}%
\bibitem [{\citenamefont {Yu}\ \emph {et~al.}(2010)\citenamefont {Yu},
  \citenamefont {Liu}, \citenamefont {Li},\ and\ \citenamefont {Oh}}]{Yu2010}%
  \BibitemOpen
  \bibfield  {author} {\bibinfo {author} {\bibfnamefont {S.}~\bibnamefont
  {Yu}}, \bibinfo {author} {\bibfnamefont {N.-L.}\ \bibnamefont {Liu}},
  \bibinfo {author} {\bibfnamefont {L.}~\bibnamefont {Li}},\ and\ \bibinfo
  {author} {\bibfnamefont {C.~H.}\ \bibnamefont {Oh}},\ }\bibfield  {title}
  {\bibinfo {title} {Joint measurement of two unsharp observables of a qubit},\
  }\href {https://doi.org/10.1103/PhysRevA.81.062116} {\bibfield  {journal}
  {\bibinfo  {journal} {Phys. Rev. A}\ }\textbf {\bibinfo {volume} {81}},\
  \bibinfo {pages} {062116} (\bibinfo {year} {2010})}\BibitemShut {NoStop}%
\bibitem [{\citenamefont {Uola}\ \emph {et~al.}(2016)\citenamefont {Uola},
  \citenamefont {Luoma}, \citenamefont {Moroder},\ and\ \citenamefont
  {Heinosaari}}]{PhysRevA2016}%
  \BibitemOpen
  \bibfield  {author} {\bibinfo {author} {\bibfnamefont {R.}~\bibnamefont
  {Uola}}, \bibinfo {author} {\bibfnamefont {K.}~\bibnamefont {Luoma}},
  \bibinfo {author} {\bibfnamefont {T.}~\bibnamefont {Moroder}},\ and\ \bibinfo
  {author} {\bibfnamefont {T.}~\bibnamefont {Heinosaari}},\ }\bibfield  {title}
  {\bibinfo {title} {Adaptive strategy for joint measurements},\ }\href
  {https://doi.org/10.1103/PhysRevA.94.022109} {\bibfield  {journal} {\bibinfo
  {journal} {Phys. Rev. A}\ }\textbf {\bibinfo {volume} {94}},\ \bibinfo
  {pages} {022109} (\bibinfo {year} {2016})}\BibitemShut {NoStop}%
\bibitem [{\citenamefont {Wolf}\ \emph {et~al.}(2009)\citenamefont {Wolf},
  \citenamefont {Perez-Garcia},\ and\ \citenamefont
  {Fernandez}}]{PhysRevLett2009}%
  \BibitemOpen
  \bibfield  {author} {\bibinfo {author} {\bibfnamefont {M.~M.}\ \bibnamefont
  {Wolf}}, \bibinfo {author} {\bibfnamefont {D.}~\bibnamefont {Perez-Garcia}},\
  and\ \bibinfo {author} {\bibfnamefont {C.}~\bibnamefont {Fernandez}},\
  }\bibfield  {title} {\bibinfo {title} {Measurements incompatible in quantum
  theory cannot be measured jointly in any other no-signaling theory},\ }\href
  {https://doi.org/10.1103/PhysRevLett.103.230402} {\bibfield  {journal}
  {\bibinfo  {journal} {Phys. Rev. Lett.}\ }\textbf {\bibinfo {volume} {103}},\
  \bibinfo {pages} {230402} (\bibinfo {year} {2009})}\BibitemShut {NoStop}%
\bibitem [{\citenamefont {Schneeloch}\ \emph {et~al.}(2013)\citenamefont
  {Schneeloch}, \citenamefont {Broadbent}, \citenamefont {Walborn},
  \citenamefont {Cavalcanti},\ and\ \citenamefont
  {Howell}}]{Schneeloch2013PRA}%
  \BibitemOpen
  \bibfield  {author} {\bibinfo {author} {\bibfnamefont {J.}~\bibnamefont
  {Schneeloch}}, \bibinfo {author} {\bibfnamefont {C.~J.}\ \bibnamefont
  {Broadbent}}, \bibinfo {author} {\bibfnamefont {S.~P.}\ \bibnamefont
  {Walborn}}, \bibinfo {author} {\bibfnamefont {E.~G.}\ \bibnamefont
  {Cavalcanti}},\ and\ \bibinfo {author} {\bibfnamefont {J.~C.}\ \bibnamefont
  {Howell}},\ }\bibfield  {title} {\bibinfo {title}
  {Einstein-{P}odolsky-{R}osen steering inequalities from entropic uncertainty
  relations},\ }\href {https://doi.org/10.1103/PhysRevA.87.062103} {\bibfield
  {journal} {\bibinfo  {journal} {Phys. Rev. A}\ }\textbf {\bibinfo {volume}
  {87}},\ \bibinfo {pages} {062103} (\bibinfo {year} {2013})}\BibitemShut
  {NoStop}%
\bibitem [{\citenamefont {Yu}\ \emph {et~al.}(2022)\citenamefont {Yu},
  \citenamefont {Mao}, \citenamefont {Niu}, \citenamefont {Chen}, \citenamefont
  {Li},\ and\ \citenamefont {Fan}}]{ysx2022}%
  \BibitemOpen
  \bibfield  {author} {\bibinfo {author} {\bibfnamefont {S.}~\bibnamefont
  {Yu}}, \bibinfo {author} {\bibfnamefont {Y.-L.}\ \bibnamefont {Mao}},
  \bibinfo {author} {\bibfnamefont {C.}~\bibnamefont {Niu}}, \bibinfo {author}
  {\bibfnamefont {H.}~\bibnamefont {Chen}}, \bibinfo {author} {\bibfnamefont
  {Z.-D.}\ \bibnamefont {Li}},\ and\ \bibinfo {author} {\bibfnamefont
  {J.}~\bibnamefont {Fan}},\ }\bibfield  {title} {\bibinfo {title} {Measurement
  uncertainty relation for three observables},\ }\href
  {https://arxiv.org/abs/2211.09816v1} {\bibfield  {journal} {\bibinfo
  {journal} {arXiv preprint:2022.09816}\ } (\bibinfo {year}
  {2022})}\BibitemShut {NoStop}%
\bibitem [{\citenamefont {Li}\ \emph {et~al.}(2022)\citenamefont {Li},
  \citenamefont {Mao}, \citenamefont {Weilenmann}, \citenamefont {Tavakoli},
  \citenamefont {Chen}, \citenamefont {Feng}, \citenamefont {Yang},
  \citenamefont {Renou}, \citenamefont {Trillo}, \citenamefont {Le},
  \citenamefont {Gisin}, \citenamefont {Ac\'{\i}n}, \citenamefont
  {Navascu\'es}, \citenamefont {Wang},\ and\ \citenamefont {Fan}}]{Li2022PRL}%
  \BibitemOpen
  \bibfield  {author} {\bibinfo {author} {\bibfnamefont {Z.-D.}\ \bibnamefont
  {Li}}, \bibinfo {author} {\bibfnamefont {Y.-L.}\ \bibnamefont {Mao}},
  \bibinfo {author} {\bibfnamefont {M.}~\bibnamefont {Weilenmann}}, \bibinfo
  {author} {\bibfnamefont {A.}~\bibnamefont {Tavakoli}}, \bibinfo {author}
  {\bibfnamefont {H.}~\bibnamefont {Chen}}, \bibinfo {author} {\bibfnamefont
  {L.}~\bibnamefont {Feng}}, \bibinfo {author} {\bibfnamefont {S.-J.}\
  \bibnamefont {Yang}}, \bibinfo {author} {\bibfnamefont {M.-O.}\ \bibnamefont
  {Renou}}, \bibinfo {author} {\bibfnamefont {D.}~\bibnamefont {Trillo}},
  \bibinfo {author} {\bibfnamefont {T.~P.}\ \bibnamefont {Le}}, \bibinfo
  {author} {\bibfnamefont {N.}~\bibnamefont {Gisin}}, \bibinfo {author}
  {\bibfnamefont {A.}~\bibnamefont {Ac\'{\i}n}}, \bibinfo {author}
  {\bibfnamefont {M.}~\bibnamefont {Navascu\'es}}, \bibinfo {author}
  {\bibfnamefont {Z.}~\bibnamefont {Wang}},\ and\ \bibinfo {author}
  {\bibfnamefont {J.}~\bibnamefont {Fan}},\ }\bibfield  {title} {\bibinfo
  {title} {Testing real quantum theory in an optical quantum network},\ }\href
  {https://doi.org/10.1103/PhysRevLett.128.040402} {\bibfield  {journal}
  {\bibinfo  {journal} {Phys. Rev. Lett.}\ }\textbf {\bibinfo {volume} {128}},\
  \bibinfo {pages} {040402} (\bibinfo {year} {2022})}\BibitemShut {NoStop}%
\bibitem [{\citenamefont {Feynman}(1967)}]{Feynman}%
  \BibitemOpen
  \bibfield  {author} {\bibinfo {author} {\bibfnamefont {R.}~\bibnamefont
  {Feynman}},\ }\href
  {https://mitpress.mit.edu/9780262560030/the-character-of-physical-law/}
  {\emph {\bibinfo {title} {The character of physical laws}}}\ (\bibinfo
  {publisher} {MIT Press},\ \bibinfo {year} {1967})\BibitemShut {NoStop}%
\bibitem [{\citenamefont {Barchielli}\ \emph
  {et~al.}(2018{\natexlab{b}})\citenamefont {Barchielli}, \citenamefont
  {Gregoratti},\ and\ \citenamefont {Toigo}}]{Toigo2018}%
  \BibitemOpen
  \bibfield  {author} {\bibinfo {author} {\bibfnamefont {A.}~\bibnamefont
  {Barchielli}}, \bibinfo {author} {\bibfnamefont {M.}~\bibnamefont
  {Gregoratti}},\ and\ \bibinfo {author} {\bibfnamefont {A.}~\bibnamefont
  {Toigo}},\ }\bibfield  {title} {\bibinfo {title} {Measurement uncertainty
  relations for discrete observables: Relative entropy formulation},\ }\href
  {https://doi.org/10.1007/s00220-017-3075-7} {\bibfield  {journal} {\bibinfo
  {journal} {Communications in Mathematical Physics}\ }\textbf {\bibinfo
  {volume} {357}},\ \bibinfo {pages} {1253} (\bibinfo {year}
  {2018}{\natexlab{b}})}\BibitemShut {NoStop}%
\bibitem [{\citenamefont {Nguyen}\ \emph {et~al.}(2020)\citenamefont {Nguyen},
  \citenamefont {Designolle}, \citenamefont {Barakat},\ and\ \citenamefont
  {G{\"u}hne}}]{Guhne2020}%
  \BibitemOpen
  \bibfield  {author} {\bibinfo {author} {\bibfnamefont {H.~C.}\ \bibnamefont
  {Nguyen}}, \bibinfo {author} {\bibfnamefont {S.}~\bibnamefont {Designolle}},
  \bibinfo {author} {\bibfnamefont {M.}~\bibnamefont {Barakat}},\ and\ \bibinfo
  {author} {\bibfnamefont {O.}~\bibnamefont {G{\"u}hne}},\ }\bibfield  {title}
  {\bibinfo {title} {Symmetries between measurements in quantum mechanics},\
  }\href {https://doi.org/10.48550/arXiv.2003.12553} {\bibfield  {journal}
  {\bibinfo  {journal} {arXiv preprint:2003.12553}\ } (\bibinfo {year}
  {2020})}\BibitemShut {NoStop}%
\end{thebibliography}%

\end{document}